# Mikhail Vasil'evich Lomonosov

## *Discourse on Greater Accuracy of Navigation* [1]

**read in public meeting of the St.Petersburg Imperial Academy of Sciences on the 8th day of May, 1759, by Collegiate Councilor and Professor Mikhail Lomonosov**

[ *Meditationes de via navis in mari certius determinanda praelectae in publico conventu academiae Scientiarum Imperialis Petropolitanae Die VIII. Mai, A.A. 1759 Auctore Michaele Lomonosow Consilario Academico* ]

[ *Рассуждение о большей точности морского пути* ]

### PREAMBLE

Reckoning the benefits accrued by humanity via seafaring human family is the same as going into an immeasurable abyss, listeners. From the most ancient times to our centuries clear testimonies about that are given by populous seaborne trade between great nations and exchange of mutual abundances. In our times, distant sea voyages to the Indian and American shores manifest additional evidences in that! Since the times when heroic efforts of Portuguese and Spaniards unlocked the untested ocean and opened it to other European nations, the fleets have greatly grown and spreading everywhere, multiplied the treasures and powers of the sovereigns and profits of the subjects. Glorious advances happened for the Europeans who through navigation aiming for the acquisition of wealth became able to reach from the ascending to the setting sun. However, it often happens that tragedies of a long way outweigh the joys from expected profits and, moreover, sometimes the hope of acquiring dies together with life. Dusting by ferocious sea, suffering from heat, thirst, hunger, fever, getting infected with a pestilent ulcer, or, worse, rabies, and meanwhile not being able to find known harbor for refuge and rest, - is nothing more than a living to lie in a coffin. All these calamities occur from deficiency of navigation, which for that has been improving with worthy diligence since the most ancient times. Nowadays, due to efforts of people ingenious in astronomy and in navigation, it progressed so far that many of the difficulties which appeared insurmountable, are now overcome and explained, and we applaud and use them with benefit in action. An important reason was that different countries established great awards[2] and thus, brought the attention of sciences and crafts to the issue. So,

---

[1] Translated to English from the Russian version of Ref. [1] by V.Shiltsev; footnotes and commentary by V.Shiltsev
[2] Here, Lomonosov apparently references several incentives in the form of large monetary prizes, established by various states for the development primarily of a method for determining a ship's geographic longitude at sea. Thus, Philip III of Spain announced a reward of 1000 crowns for whoever develops a method for determining longitude at sea. Subsequently, with the same goal, a prize of 10,000 florins was established by the Dutch States-General. The issue of determining longitude from measurements of so-called lunar distances occupied the minds of sailors and astronomers for over two centuries (until the invention of the chronometer, after which the need to

although my work might seem useless attempt to add something beyond these works, but I followed the mine prospectors, who are nourished by sweet hope while sometimes without any probability and not always in vain. Thus, dismissing any hesitations, I present everything that I reflected, invented and produced on this matter.

As is known, there are two different methods to find and determine position of a ship at sea. The first is to find the latitude from heights of the stars and the longitude by comparison of the time at the ship's meridian with that at the zero meridian. The second is to calculate the longitude and latitude of the ship from the compass direction and the ship's speed, which is either measured with lug or estimated from the records the wind strength and the number and position of the sails.

The first method is only applicable in clear weather, while the second can be used at any time. Despite the numerous difficulties associated with both methods, those who have sought ways to mitigate them have demonstrated their ingenuity and strength, and some have even attempted to put them into action. Each difficulty will be briefly presented here to clearly show the order of this discussion and my efforts in this matter.

In clear weather, observations typically begin by determining the latitude of the location based on the elevation of celestial bodies above the horizon. Then, the time is deduced from the different elevations of two or one body at the same time. Nowadays, these observations are effectively conducted using the English quadrant[3] with mirrors, an ingenious invention that enables the alignment of stars in the sky. After determining the latitude and time at the ship's location, the longitude is sought through two different methods: mechanical and astronomical. The former involves comparing the positions of stars, while the latter involves determining the difference in meridians using the most reliable marine chronometers available.

The inconveniences and difficulties associated with this method include the following: although the English Hadley quadrant is highly capable of measuring the altitude of stars above the horizon, it does not eliminate the ship's motion, which extends directly from the star to the

---

solve this problem disappeared) and could not be solved because the fundamentals of the theory of the moon's motion remained undeveloped until Newton's time. It follows that the pre-calculation of more or less accurate lunar ephemerides, necessary for sailors to determine longitude, was still impossible at that time. The positions of many other celestial bodies were also not determined accurately enough by that time. All this served as the reason for the establishment in 1675 of a special astronomical observatory in Greenwich. Finally, with the aim of encouraging astronomers and sailors, a permanent "Commission for the Discovery of the Longitude at Sea" ("Board of Longitude") was established in England in 1714 and existed until 1828, which had significant funds for annual rewards (up to 2000 pounds) announced for the development of the specified methods, as well as for individual discoveries and inventions that could be applied in navigation.

[3] The quadrant is an angular instrument, invented in 1731 by Hadley, who called it an octant because its arc equaled 1/8 of a circle. However, with the use of the mirror system, the instrument allowed measuring angles up to 90°, i.e., within ¼ of a circle, which is why it was also called a quadrant. Subsequently, the instrument began to be made with a longer arc (1/6 part of a circle), and then it was called a sextant.

observer. Additionally, the instrument does not mitigate the perpendicular oscillations to the observer, resulting in difficulty accurately determining the distance of stars from the horizon. Furthermore, variations in the height of the horizon due to different rays of refraction and the indistinct limit during nighttime or foggy conditions lead to significant errors in all observations, with errors barely less than five minutes even in clear weather. Consequently, discrepancies in latitude and time can lead to significant discrepancies in longitude, leaving the ship's position uncertain. Therefore, I endeavored to find a more reliable method that could be used more frequently, leaving behind the unreliable and unclear horizon.

To determine the time at the prime meridian[4], the most preferred method is considered to be marine clocks, which, through comparisons of star positions, are preferred over celestial observations. These clocks, designed to maintain accuracy over long periods of time, are affected by the turbulent sea if they employ weights, while those propelled by springs are generally preferred – and justly so. The progress made in this field in Great Britain, where these instruments are meticulously crafted to meet the highest standards, is not yet widely known[5]. However, I am not prohibited from presenting my idea to the scientific community, despite its possible insufficiency compared to the aforementioned efforts.

Moreover, the method that involves comparing the positions of stars to determine longitude at sea should not be neglected, as it possesses certain advantages over the mechanical method. Although marine clocks endowed with the necessary qualities can correct themselves without the need for tedious star observations, they are still susceptible to fragility, instability, and inaccurate rotation. Conversely, the consistent movements of celestial bodies can undoubtedly establish the desired time, provided their positions are determined accurately and consistently according to the true theory through frequent and precise observations, free from errors. Desired marine clocks may not be made by every master craftsman, nor can they be acquired by every enthusiast due to their rarity and high price. However, the instruments required for observing celestial bodies can be made more conveniently and affordably, especially those described below. Although marine clocks continuously indicate every moment of time, the positions of stars are not always visible for observation, especially when planets are in close proximity to the Sun's rays. However, this drawback, which rarely occurs, can be compensated for by a multitude of observations that not only mutually correct each other, increasing the likelihood, but also reveal the errors of the clocks themselves. Nevertheless, this matter will become clearer in its proper context.

---

[4] "Prime meridian" is the initial meridian (from which longitude is counted), which is now called the zero meridian.
[5] This refers to the invention of the chronometer. The first prototype of it was made by Harrison in 1735, and the next several chronometers were made only by 1761, and "The principles of Mr. Harrison's timekeeper" were published only in 1767. Therefore, by 1759, when this "Discourse…" was written, Lomonosov could have only heard rumors about the invention of the chronometer in England. Consequently, even before the appearance of the English chronometer in Russia and its description, Lomonosov independently conceived the idea of constructing a chronometer consisting of two drums — cylindrical and conical, with an unwinding chain, thanks to which the spring unwinds evenly. Lomonosov used this idea in the clock mechanism, which rotated the drum of the self-recording compass (see Fig. XIV below). The fundamental scheme of such a design has been preserved in later centuries' chronometers.

But now gloomy weather sets in, obscuring the Sun, Moon, and stars from view. Astronomical instruments become useless, without which even the most precise and expertly crafted clocks are of no use. Meanwhile, the storm swiftly propels the ship, waves deflect it from its intended course, the path is hastened by the capable current, and opposed by the adverse one. After enduring such conditions for several weeks, how can a sailor know where to seek refuge, where to steer clear of shallows, rocks, and shores, or where to find shelter from the steep cliffs? Therefore, sailors must seek ways to overcome these difficulties, unfortunately, there are few decent methods devised, and even fewer put into practice, although it seems that they are more necessary than the former, as storms rage fiercer in gloomy weather, and disasters loom closer. Considering this, I endeavored to devise new routes to evade such inconveniences, and, as it seems, I did not entirely fail.

I have considered two methods: the first requires instruments designed according to theory by skilled craftsmen, which, when used in advance for assurance through experiments in actual practice, can be effective. The main ones among these are: the self-recording compass[6], the dromometer-log[7], the clizeometer-driftgraph[8], the cymameter[9], and the salonometer[10], which are described in their proper place, along with an explanation of their use.

The second method requires the long-term expertise, ingenuity, and vigilance of seafarers and the constant attention of physicists and mathematicians. It mainly consists of the true theory of ocean currents and magnetic needle variations, and it should be based on accurate observations. Therefore, in the third part, I will propose the study of marine science, which I recommend to all who engage in it, with the advice of Pliny: "Countless multitudes sail the open seas to foreign shores, but for profit, not for knowledge, but a non-blinded or with a mind focused on pleasure does one contemplate that knowledge can make the profit safer."[11]

PART ONE

---

[6] Self-recording compass (see §§ 38, 39) — Lomonosov anticipated his era by two centuries with the idea of such a compass, which found wide application only in modern instrument-making and navigation technology and in principle does not differ from the modern course recorder.

[7] Dromometer (see §§ 42, 43) — a bottom mechanical log of the spinner type, another Lomonosov's invention that preceded its widespread use by almost a century. Such a type of log, mounted in the bottom of a ship, first proposed by Lomonosov, became widespread only in the 19th century.

[8] Clizeometer (see § 40) — an instrument for determining the leeway of a ship under the action of the wind, with a self-recording mechanism (driftograph); first proposed by Lomonosov.

[9] Cymatometer (see §§ 44-47) — longitudinal clinometer, an instrument with a mechanical counter of the ship's longitudinal oscillatory movements, designed to account for the influence of keel roll on its course.

[10] Salometer (see §§ 48-53) — an instrument for determining the direction and speed of currents. In later ages, a device of this kind, proposed by Lomonosov, was called "Mitchell's float".

[11] A quote translated from "Natural History" by Pliny the Elder (Historiae Naturalis, lib. II, § 118).

# ON DETERMINING LONGITUDE AND LATITUDE IN CLEAR WEATHER

## CHAPTER I
## ON DETERMINING TIME AT THE SHIP'S MERIDIAN

### § 1

In clear weather during the day, the Sun, and at night, the stationary stars, are usually used for determining latitude and time. As for daytime observations, the visible horizon is often clear, especially when the side where the Sun appears is clear and the sea surface is rippled by waves. However, the variability of ray refraction renders it unreliable, especially because the ray from the Sun traverses only a part of the atmosphere, while the one emanating from a star penetrates it entirely, making it almost impossible to reconcile the variable refractions with accurate rules. Nevertheless, latitude can be determined by this common method, and the results are satisfactory for use, which we will demonstrate shortly.

### § 2

At night, in addition to its variability, the horizon is not clear or accurate due to darkness. Therefore, I decided to determine the time at the ship's meridian more precisely from the positions of the stationary stars. It often happens that stationary stars come onto the same vertical line at the same moment, which, observed accurately regardless of the darkness and the instability of the horizon, precisely indicates the time at the ship's meridian. Similarly, it often occurs that stars appear at the same height, from which the aforementioned position can also be inferred. However, as the second method is much more convenient for calculation than the first, every effort is made to explain it.

### § 3

The instrument for observing stars on the same vertical lines that I have devised is as follows (Fig. I). An equilibrium device should be made of copper strips in the form of elongated quadrilaterals, somewhat differently than compasses are placed in cases to mitigate oscillations from waves. However, the triple *a, b, c* is arranged so that the opposing sides, freely moving around the axes *dd, ee*, tend to maintain a parallel position with the horizon. This is done to avoid mirror inclinations, as perpendicular ones are neutralized by their arrangement. Although *aa* will follow the inclinations of the ship, *bb* will remain much calmer, and *dd* will barely sense any rocking, staying in a parallel position with the horizon. In the elongated inner quadrilateral, two strips *h* and *l* are attached at an equal distance from the axes on both sides; between them, two flat metal mirrors are installed. One, *N*, is fixed at a 45-degree angle to the plane of the quadrilateral and attached; the other, *P*, rotates around the axes *rs* (Fig. I, II). A visual astronomical tube *TT* (Fig. III) of such size that it can be used without significant discomfort is attached to these mirrors. To adjust mirror *P* to different positions, as to bring stars into alignment by deflecting the ray to the same height, an endless screw *k* is used.

§ 4

Observation of two stars on the same vertical circle should be done as follows: mirror *P* is placed with another mirror *N* in the position required by the angle, the measure of which is the arc connecting the two observed stars, which should be sought in deliberately compiled tables. The angle can be widened and narrowed as much as necessary with the endless screw. With this instrument directed at the stars at the moment when they approach the same vertical circle[12], you will see them at the same elevation. And as soon as they come so close to each other that they almost converge at one point, at that time, on the ship's chronometers, or (if you subsequently decide to investigate the time difference at the prime meridian through astronomical observations) on pocket watches with seconds, you should set the time upon the stars' conjunction. If the ship's motion, despite the equilibrium of the described instrument and the ship's observatory, causes lateral mirror oscillation, resulting in the stars appearing to meet and diverge by horizontal motion, then it should be noted when the moving star in the mirror touches the star outside the mirror on one side, then, after several swings, touches it again for the last time. The time limited by these two extreme touches should be divided into two equal parts and added to the time of the first, by which the true time of the stars' position on one vertical circle will appear.

§ 5

I attempted to employ Gadleev's quadrant for such observations, which I, by my addition, dub double, for the horizontal and vertical alignment of stars, which should be briefly explained here. The large mirror, usually attached perpendicular to rule *RR* [Fig. XVII], and moving with it along arc *BB*, aligns the stars to the horizon by known angles. It should be soldered to axis *A* in such a way that its rotation brings the stars onto one vertical line, that is, upon the mirror's rotation around axis *A*, star *r* (Fig. V) will reach angle *t* at the top. According to this arrangement, as required by rule *RR*, star *r* will descend from point *t* to star s, and the observer's comrade will mark the time on the clock according to this sign, dividing the different elevations of stars *r* and *s* from the horizon on the arc. Finally, the time at which the observed stars at this latitude should rotate in the shown difference in height can be calculated.

§ 6

---

[12] On the same vertical line

Lateral oscillations of the stars brought together in one place, as shown now, cause them to shake, which, by paying attention to the observation of the first mutual star contact, can also be determined by dividing the time in half and adding half to the first or subtracting from the last contact to find the hour and so forth on the ship's meridian.

§ 7

Although, using the first instrument[13], one or the other oscillation in the first star encounter and the final parting will escape the observer's notice, any oscillation in the internal quadrilateral, and consequently in the mirrors, must last less than a second. Therefore, the error in time cannot be more than four seconds, as I hope, even in strong oscillation. The ship's tremors, which threaten its immersion and knock the instrument from the hands of the observer, depriving him of hope from the heart, will not allow even the most crudest observations.

§ 8

To reduce the monotony of precise division of the entire quadrant and to achieve greater accuracy, I consider the following method the best: 1) Divide the arc into 9 equal parts with all possible calculation. Attach to it a copper plate *LL* (Fig. V), divided into 10 degrees and each degree into 6 parts, 10 minutes each, in such a way that the division of ten degrees corresponds to the ninth part of the quadrant with the greatest possible accuracy. The moving plate shown along arc *BB* should be fixed against each ten degrees with round nails *cc*. From this follows: 1) that according to a well-known general law in mathematics, the same thing equals itself in magnitude, and the same division of each ten degrees cannot be more equally divided; 2) labor and calculation for precise division into ten degrees can be more conveniently used than for ninety. Then attach rule *RR* so that it can move along plate *LL* with endless screw *C* and wheels *SS*, by which the position of line *q,* drawn from center *C*, divided into seconds by Nonius instruction[14], can be seen, facilitated by microscope *M*, which consists of a cylinder part cut parallel to its axis and magnifies the smallest parts in width and presents them clearly to the eye.

§ 9

I use metallic mirrors and recommend using those instead of the glass ones, in which the fourfold refraction of rays, the fourfold passage through mirror glasses makes things worse: because the first usually brings the parallel position of rays into confusion, and the second dulls the strength of light. And although making flat metal mirrors is considered more difficult and expensive, I argue against it, because from one metallic mirror, twenty mirrors of half a square foot can be prepared for the aforementioned use, cut and cast in one go. The edges of the whole convexity should be feared; the middle always remains the most precise plane.

---

[13] a clinometer with an artificial horizon, described in §§ 3 and 4.
[14] more correctly - by Vernier's method. Unlike Nonius, who developed a cumbersome method of measuring arcs in angular measure, Vernier provided a simple method for measuring linear quantities, also applicable to measuring arcs.

§ 10

All this at night, when the movement of stars appears as a spectacle for this maritime use, but during the day, the various heights of the Sun from the horizon should be used in the usual way, if the help of nocturnal luminaries is hindered by expectant doubtful weather. Hadley's quadrant will provide assistance to the observer sitting in the maritime observatory (Fig. XXIII). Refraction of rays, extending from luminaries and the horizon, as mentioned above (§ I), should be somewhat corrected by a theory of refractions composed from observations, based on the following: if the amount of refraction corresponds to the amount of transparent matter, that is, in this case, air, then, of course, the quantity pierced by a ray is the measure of refraction[15]. Therefore, the quantity of air lying on the visible horizon corresponds to the height of the barometer, so that the higher the mercury stands, the more refraction of rays there should be[16]. With many observations of stars and comparing their refractions with the height of the barometer, this task may be considered achievable over time.

§ 11

By observing the fixed stars at night on one vertical circle[17], the time on the ship's meridian can be determined by the following methods: 1) if the stars are on the same meridian, which rarely happens, then the calculation is very easy: because the degrees enclosed between the vertical circle and the equinox collure[18] show the time without knowing the latitude; 2) when the stars observed on one vertical circle are not on the same meridian, one should first select a star close to the pole, such as the North Pole Star or others composing the constellation of the Little Dipper. This is done so that, having first ascertained, though not precisely, the latitude by the usual method, the time can be determined in the following manner.

§ 12

---

[15] Here Lomonosov formulates the basis of the theory of average astronomical and terrestrial refraction, the phenomenon of which, without a scientific explanation, had been noted earlier (in 1559 by E. Wright). Based on this theory, the formula for the average astronomical refraction was subsequently derived: $r = 58''\,\text{ctg}\,h'$, where $h'$ is the observed altitude of the luminary object (star).

[16] Another important point formulated by Lomonosov, refining the law of terrestrial refraction; it takes into account changes in the magnitude of terrestrial (and astronomical) refraction due to barometric pressure. It was later established that the magnitude of refraction also depends on changes in air temperature.

[17] Here – the meridian.

[18] Nowadays called "colure," i.e., the circle of declination of the point of the vernal equinox.

Let the north pole be *P* (Fig. VII), the zenith *Z, D* - the Pole Star, *F* - a star in the observation of its comrade. Let line *ZD* be a vertical arc, *PF* - an arc of the ship's meridian, *PF* - an arc between the pole and its comrade, *DP* - between the pole and the Pole Star; all arcs of the largest circles, from which *PD* and *PF*, by the declination of the Pole Star and its comrade, and *FD* - by angle *N*, are known; thus, the whole triangle *PFD* will be found by spherical rules. And by the known elevation of the pole, line *ZP* is known; thus, from the given arcs *ZP* and *FP* and the angle at side a at angle t, the other parts of triangle *FPZ* are found. Finally, the found angle *b* must be added or subtracted from the angle between the first meridian *mP* and the line *FP*: the sum or the remains will be the difference between the first meridian *mP* and the ship's meridian *ZP* and the measure of time by the passage of the equatorial circle through the ship's meridian.

### § 13

The accuracy of the latitude is less required the closer the observed stars are to one meridian and the sharper the angle between *ZP* and *ZD*. For this purpose, the Pole Star is all the more capable; another star may be lower than the pole in the case of its great elevation in northern countries.

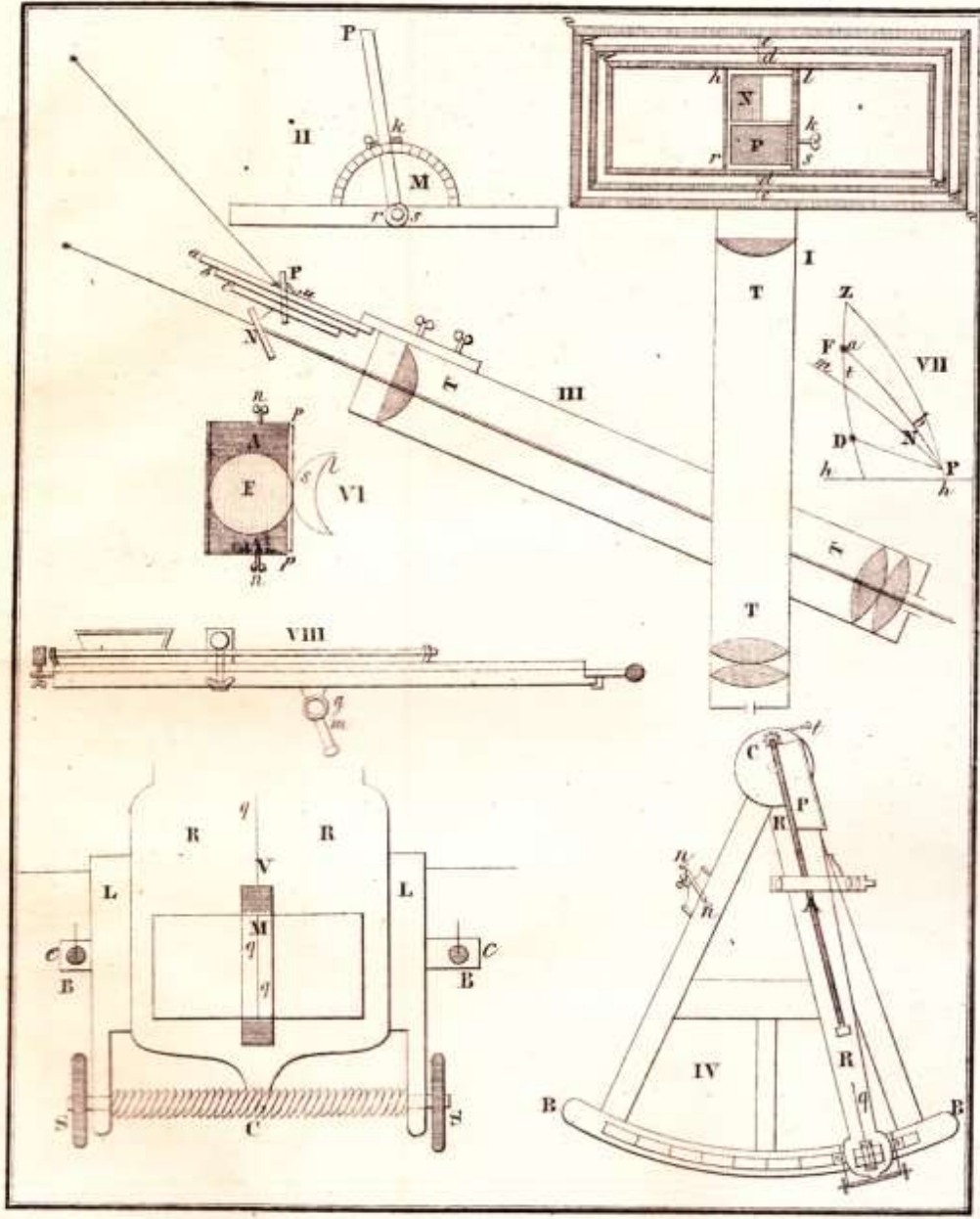

Figs. I-VIII.

# CHAPTER II
## ON FINDING THE SHIP'S LATITUDE BY THE DETERMINED TIME

### § 14

Although the latitude determined by ordinary observations is considered sufficient at sea, because the error is about five or six minutes, which is considered insignificant, and for the method proposed by me to determine the time accurately, however, in my opinion, the latitude more precisely determined is not only useful for sailors themselves, but also gives much assistance for checking other methods proposed in the second part. Therefore, especially in this chapter, I show how, leaving aside the horizon, a much more precise latitude can be obtained from the determined time.

### § 15

This should be sought in a slightly different way from that by which the time is found (§ 12) on the ship's meridian. Two stars on the same vertical circle should be observed with the instrument and method shown above, especially those which quickly flow through the mentioned line, meeting, such as those which differ from each other by longitude and latitude. Many can benefit greatly from these in clear weather, choosing any pair, anyone who has only moderate knowledge of astronomy.

### § 16

From the observation, it is evident that the line extending from *Z* (Fig. VII) through *FD* to *hh* is vertical. Lines *PF* and *PD* from the pole to the observed stars are arcs of the largest circles; likewise, the angle between them to the pole is known from the schedule of fixed stars; therefore, each part of triangle *RFD* is known by spherical trigonometry. Then the distance of the circle *Pm* from the ship's meridian *ZP* is found by the definition of time (§ 12) on the same meridian; hence the angle *mPZ* is known. But since the angle *FPm* is also known by the distance of the circle from the arc *PF* from the star catalog, subtract it from the angle *mPZ*; the remainder will be angle *b*. Finally, the side angle *a* from the known angle *PFD* or *t* is known; then the two angles *a* and *b* and the arc *PF* in the triangle *ZPF* will be known, from which, among other things, the arc *ZP*, as the complement to the arc*Ph*, that is, the elevation of the pole at the ship's location, will be determined.

### § 17

It is already quite clear that observations for determining time and latitude on the ship's location without using the horizon according to the prescribed rules at night can be used when a large number of stars almost continuously present themselves for this purpose, so that by repeating observations as many times as necessary, the time and latitude of the place can be determined with extreme accuracy.

# CHAPTER III
## ON INDICATING TIME ON THE PRIME MERIDIAN WITH CLOCKS

### § 18

Clocks swung by plumb lines and moved by weights are by no means suitable for indicating time amidst the swaying of a ship at sea. Those operated by springs can be effectively used in the following manner: spring-driven clocks (the larger, the more reliable they can be made) with seconds, and to avoid stopping when wound, arrange them in one box so that they can be wound at different times; for example, let the winding of the first clock begin at noon, the second at the end of the sixth hour in the afternoon, the third at midnight, the fourth at six o'clock in the morning (in large clocks, quarters of the day can turn into whole days). In this way, errors arising from the unevenness of the spring forces and other parts composing the clocks can be largely avoided, for the sum of the times shown on different clocks, divided by four, will divide the errors, which, by canceling each other out, will come closer to the true time.

### § 19

Clockmakers' ingenuity can bring together four springs and as many spiral ones to move a single wheel and to employ their forces and accuracies on it, and by which other structure of the clock can be controlled by one pendulum. *E* (Fig. XIV) represents the springs; *c* - the spiral ones; *A* - the wheel onto which the combined forces are exerted; *t* - the gear by which the whole structure of the clock revolves. In my opinion, the pendulum should be a solid circle, cut from strips such as those used for minting coins and on which one can rely for density and uniform thickness.

### § 20

These clocks can be protected from the ship's motion and from changes in temperature and cold in the following manner. Firstly, boxes suspended on wire-coiled springs do not feel sudden impacts so much, to which ordinary compass balancers can add much calmness. Changes arising from temperature and cold should be avoided in the following way. Place the clocks inside the ship, in a part submerged in the sea, where the air's dissolution changes little. Moreover, this position, being in the middle of the ship, is less subject to fluctuations. Small pocket watches, corrected for such immovable clocks, should be set up and used during observations.

### § 21

But all these difficulties can be more easily avoided, it seems, if as much effort is put into sand clocks (called hourglasses), using metallic particles, as into spring-driven ones, about which I reason as follows. First, stretch a thin wire into hair, then cut it into short pieces so that they are equal in length and thickness, cut by special scissors, and can be cut many at once. Mix this material with a sufficient amount of ground charcoal, put it in a pot in a melting furnace so that all the particles melt into balls by intense heat of fire, then wash them, polish them with a brush. Such small balls should serve much better than sand for hourglasses, because they are smooth, uniform, heavier than sand, and in a word, are a liquid substance, having no connection between the particles and a surface free from vibration.

§ 22

Then, in glass jars connected in the usual way, instead of perforated tin, put steel cones on both sides of the hole, like funnels, so that copper sand (or, even better, silver) can be poured continuously in both directions. Finally, the amount of metallic sand should be measured experimentally with accurate wall clocks so that the end of the flow accurately determines one hour or more.

§ 23

Such metallic hourglasses are not afraid of changes in temperature and cold or thickening of oil used for their free movement. Forced movements, as from spring clocks, can also be avoided. How much the flow of metallic particles or sand can be accelerated by swaying should be investigated by art, to know how much to add or subtract in comparison with time.

§ 24

The use of these pouring clocks differs greatly from spring clocks, for after the expiration of the metallic grains, they must be turned over, counting one second for what should be considered one hour. And if they are made for one hour, each turn of the clock should mean something, for which it is necessary to attach a wheel to the axis, divided into parts. For after turning the clocks at the end of the flow, minutes and seconds must be counted by pocket watches, which can go for one hour without error, and by them to make astronomical observations on the ship's meridian, comparing them with the time of the prime meridian, and deduce the longitude of the place.

# CHAPTER IV
## ON DETERMINING THE PRIME MERIDIAN BY OBSERVATION OF STARS

§ 25

Observations of the distances of the Moon from fixed stars are highly regarded in determining the time on the prime meridian. It is necessary to consider this method above all others, for although star occultations may seem much more precise than measuring distances, it rarely happens, and observations cannot be arbitrarily undertaken to determine the position of the Moon more accurately. Meanwhile, I make efforts to show that observation and measurement of distances, by which the stars appear to be separated from the Moon, are much more convenient and accurate.

§ 26

Attach a handle *m* (Fig. VIII) to the Hadley's quadrant, which would be secured by a sphere, closely attached to another thin handle, *g*. Thus, direct the instrument in such a way that its plane is sufficiently parallel to the plane of the lunar ecliptic or to another plane containing the Moon, the star, and the observer's eye, so that all previous observations, knowing the difference in altitude between the Moon and the star in degrees, can be established. The observer, sitting in the ship's observatory and being free from great oscillations, should be able to adjust with his accustomed hand.

§ 27

The Sun's light obscures the Moon if brought close to it, and the Moon obscures the star which is brought close to it. Therefore, I sought means; what I found will suffice: that is, to the smaller mirror of the Hadley quadrant, attach with screws *nn* (Fig. VI) a thin copper strip *A*, covered with light ink, in which the image *F* of the Sun or the Moon can be clearly seen, and directly visible stars do not obscure the Moon or the Sun. Leave part of the smaller mirror, which is at the edge *pp*, open so that a very small segment *S* of the Sun or the Moon can be clearly seen and noted in conjunction with the observed star. Usually, smoked glasses are used in such cases, but here they are unsuitable, because not only does the light of the Sun or the Moon become dull on the edge, but the observed star is completely obscured, because it, being brought closer, must let its weak ray pass through the same black glass.

§ 28

In such observations, it should be noted that if, due to the rocking of the instrument, the observed star sways perpendicular to the plane of the quadrant, one should wait until it touches the unobscured part of the lunar arc at its very top for the first time, and then set the time. If parallel passages and departures occur, then the moment of the first passage should be noted, as well as the last departure beyond the mirror, divide the time in half and, adding to the first approach or subtracting from the last appearance, one can determine the moment when the luminaries will be so far apart as the degrees and parts shown by the division of the quadrant.

§ 29

From experiments conducted with the utmost precision and diligently repeated on the distance of different stars, preceding and following the Moon, one should make calculations based on lunar tables, which have been much corrected by the unremitting efforts of learned men and still require even more precise adjustments. Therefore, I argue that those who strive for great success in this matter should use an instrument for observing the distances from the Moon to fixed stars, similar to Hadley's quadrant, but larger and specially made for this purpose, with which many observations can be made in one night at a stationary observatory. For other common methods compel the astronomer to divert his attention to two points. On the contrary, by connecting the Moon with the stars, all his vision and attention can be directed to one place. The description of such a quadrant I leave for another time.

§ 30

This is how the Moon assists sailors at night. But the Sun is not without similar use during the day, when the Moon is visible on the horizon, the distance from the Sun, measured by an octant, can show the time on the prime meridian, and repeated observations from different distances at different times can serve instead of the distance of different stars from the Moon.

§ 31

Although the satellites of the highest planets cannot satisfy sailors with precise time limits, in long voyages, when sometimes knowledge of longitude with an error of two or three degrees is needed, when the new moon is not visible, they can bring considerable help, because the usual error in time is about ten minutes.

§ 32

For observing eclipses and appearances of the highest planets, an astronomical tube with a mirror can be used, attached to it as follows. Let there be a tube *TT* (Fig. XIX), at the top of it attach a handle *ss* with a compass balance *AA* and with two wheels. One upper *R* is twice the diameter of the other *r*; both move together in the cord *ff*. Set the axis of the smaller one in balance, the larger one in the handle. To this wheel, divided into degrees, attach a light mirror, which can be set with an endless screw in the desired position, as the height of the planet requires. Thus, when the tube is lowered by swaying to the horizon and raised from it, the ray coming from the star into the tube will deviate little from the axis of the tube, and the star will always be visible, for when the smaller wheel rotates, for example, 10 degrees, then the larger one will move only five, and the ray, by its deviation from the mirror, will add the same five degrees to it. And thus the required follows from the above.

# PART II
# DETERMINING LONGITUDE AND LATITUDE IN CLOUDY WEATHER

## CHAPTER I
## MANAGING A SHIP ON THE SEA SURFACE

§ 33

Everything proposed in the first part can only be utilized by sailors in clear weather. But as soon as the sky becomes covered with clouds and the stars are obscured from sight, then neither the best marine chronometers, nor astronomical instruments, nor the devices freeing the ship from oscillation can be employed. Therefore, it is evident that another refuge must be sought. It is astonishing that so little attention is paid to devising, utilizing, and refining such methods by those navigating the great seas, knowing that a significant portion of the time the sky is covered with clouds, and then the sea becomes more violent, diverting ships from their intended course and casting them into the jaws of inevitable fate.

§ 34

In such conditions of sky and sea, the general and constant guide is the compass. Its force-driven steel needle shows the way in the absence of celestial luminaries, which were the sole guides to ancient sailors. In gloomy weather, they had to keep close to the shores, especially during storms. Our inquisitive times have brought forth so much concern for understanding the compass that this lifesaving invention no longer seems as important as it used to be, if we cannot find the reasons for its variations and accurately correct them according to different places and times.

§ 35

Although we already have repeated successes in the exploration of the laws of magnetic force[19], yet due to the neglect of sailors and the entrenched habit, which obstructs progress in all sciences, attention is turned away from them. A notable example is the negligence in observing changes in the declination and inclination of the magnet, upon which salvation and destruction depend. If there were, as there should have been long ago, a sufficient number of properly conducted observations, certainly the true theory of magnetic declination and inclination would have been deduced by the reasoning of physicists.

---

[19] some deviation of the magnetic needle from the true meridian (magnetic declination) was noted by sailors as early as the 13th century. Two centuries later, it became known that declination is not uniform in different parts of the globe. This was reflected in German road maps dating from around 1492. During de Castro's journey to the East Indies in 1538-1541, magnetic declination was determined at 43 points. In 1536, the phenomenon of magnetic inclination was discovered. In 1634, it was proven that magnetic declination undergoes long-term changes (centennial variations).

§ 36

This mostly arises from the fact that sailors use small and poorly made compasses, which not only prevent accurate observations at sea but also misdirect the course of navigation.

§ 37

Compasses should be made larger so that the division of the winds is clearer, and together with degrees, the helmsman could pay attention even to the smallest 1/360th of the degrees of the compass circle. It should be positioned so that the black line drawn on white is precisely parallel to the ship's axis or keel, and the highly magnetized steel could overcome friction. This is sufficient for an ordinary compass used for navigation. However, to know all the errors that occur due to the helmsman's mistakes, he must have a special self-recording compass, which can be made as follows.

§ 38

In the same box *AA* (Fig. IX, X), alongside the compass, fit spring clocks *CC,* with which shaft *D* moves, wound around with paper *EE,* which is then wound onto another shaft *H.* The circle *BB*, on which the winds and degrees are depicted, should be mounted on a steel piece made through the Knight's method of magnet-making[20], which can overcome the slight friction of a thin pencil without delay. Direct the movement of the circle on the through axis *ii* so that at the bottom of the box and at the top to the glass it is installed and so that the circle, with both the bottom and the glass, stands parallel, and the paper, winding from one shaft to another, is perpendicular to the plane of the circle, and the diameter of the compass circle, extending along the keel, is also perpendicular. Then, near the compass circle, make a ring *mm*, which could be turned with a pencil to the side where the ship needs to be directed. The pencil should be on the lightest spring of wire, and, in short, everything should be delicate and gentle.

§ 39

By attaching clocks to the compass in this manner, the shaft will rotate, and the paper will wind onto the other; the pencil, lightly touching it, will draw a line, indicating any deviations and errors in navigation, which can generally be seen and measured by weight. It may seem strange to identify errors in navigation by weight, but it is possible. That is, deviations towards *N* (Fig. XI) from the straight line *KK*, drawn on paper, should be cut out and hung on very sensitive scales, such as those used for weighing precious metals. The weight will show which side the ship's deviation was greater, and the difference between the two weights after subtraction will be the measure of the excess deviation on one side or the other.

---

[20] Gowin Knight (1713-1772) - was a British physicist who, in 1745, discovered a process for producing strongly magnetized steel, employed for fabrication of magnetic needles for high-precision compasses, his technologically advanced compasses were adopted by the Royal Navy; a Fellow of the Royal Society, the Copley Medal (1747).

§ 40

With this, as I think, all the errors that often occur due to the negligence of the helmsman can be identified and eliminated. But there are still major inaccuracies when a side wind deviates the ship from its intended course. The angle included by the line of the ship's direction *CD* (Fig. XII) and the line along which the ship moves due to the lateral force *Kl*, I advise measuring with an instrument attached behind the cabin near the ship's axis (I call this instrument a clinometer). Attach a quadrant[21] *Q* with degrees, divided in half by the line parallel to *CD* together with the keel, with a spoke *F* and a pointer *h*, to a thin rope about forty fathoms (or longer, the better) tied to the end of the spoke, a stick *l*, which, being stretched by remaining water, will indicate the angle of deviation on the quadrant with the pointer. The oscillations of the pointer from the swells can be observed in both directions, and the midpoint can be taken as the true deviation[22]. However, if someone adds simple spring clocks to this, as above with the compass, they will have a self-recording clinometer, which by changing the wrapped paper at the appointed time will clearly show the ship's deviation in its direction from the side wind before their eyes.

§ 41

There are other ways to detect such deviations when the ship is swaying with extreme agitation, and for this, the use of a clinometer is useless, because artillery science has mixtures from which small entertaining lights on the water appear, by throwing filled tubes from the stern, night with the light of fire, and during the day with the ascent of smoke will show the deviation of the ship from the intended course.

CHAPTER II
**MEASURING THE SPEED OF A SHIP ON THE SEA SURFACE**

§ 42

Log lines, or measuring ropes of the ship's speed, do not show its changes continuously, but intermittently. Hence it is quite evident that those methods should be preferred which show this continuously. To achieve this, a device must be made that constantly moves[23], showing the speed at every moment; and so that when the course changes, the amount of distance traveled can be observed at a glance without the tedious process of letting out the log line and winding it back.

---

[21] Here it refers to a simple sector.
[22] the leeway of the ship due to drift.
[23] The "dronometer"

§ 43

Make plan *A* (Fig. XXII) spiral-shaped, which, when installed along the keel axis, would rotate around it near the water. Attach this plan to an iron strip *cc*, which can be hooked onto the keel with iron hooks *dd* from below, and can be secured and passed through the stern post with the upper end. Let a thin rope *f* run around the gear wheel, which shares a common axis with the plan, and together with the wheel *e*, which rotates other wheels, so that the revolutions, determined by skill, on the wheel *m* indicate fathoms, and on the others *hg* — miles, which all should be produced by gears.

§ 44

Meanwhile, when the ship rises and falls with the waves, the distance traveled, as shown by the designated machine, follows an uneven arc on the surface described by the ship's course through the current, forming a very curved line, that is, the one described by plan *A*. Therefore, a distance measurer without the help of another instrument, which could properly be called a "cymatometer," will not show, because it calculates the waves that rock the ship and shows all deviations from the horizon.

§ 45

Make a plumb line *A* (Fig. XX), attached to board *BB*, which should be hung parallel to the keel of the ship, so that, swinging along its length, it would also incline with the same angles, while freely rotating in the lateral directions. Attach a toothed wheel on axis *C* to the center, so that when the plumb line swings away from the center to *D*, hook *K* would grab onto the teeth of the wheel, and, returning from *D*, would pull back by as many degrees from the initial position as *A* moves away from partition *g*. Thus, the degrees of all swings would be measured with each movement of the plumb line. Wheel *N* will show the number of revolutions of wheel *C*. Thus, it will be possible to determine in a certain time how many degrees were in all swings.

§ 46

When this happens, with each touch of the plumb line to partition *g*, nail *i* is pushed in, which cannot enter the hole any further, except to grab onto one tooth of wheel *M* and, by the force of spring *e*, be forced to return, thus moving the wheel, the return of which is prevented by the latch *p*. The rotations of this wheel *M* are indicated by another wheel *N*. Thus, with this rotation, the number of sways and oscillations will be determined, and together with the above-written, the total number of degrees at one time by one instrument.

## § 47

Having the total number of degrees from all swings, it must be divided by the number of sways or ship inclinations: this will result in the total angle to the horizon. Knowing this, it can be compared to the curved line of the general waves with the arc, which is the path of the ship on the sea surface, and from there, its true distance can be found. How this calculation should be performed seems worthy for the ingenious mathematicians of our age to work on.

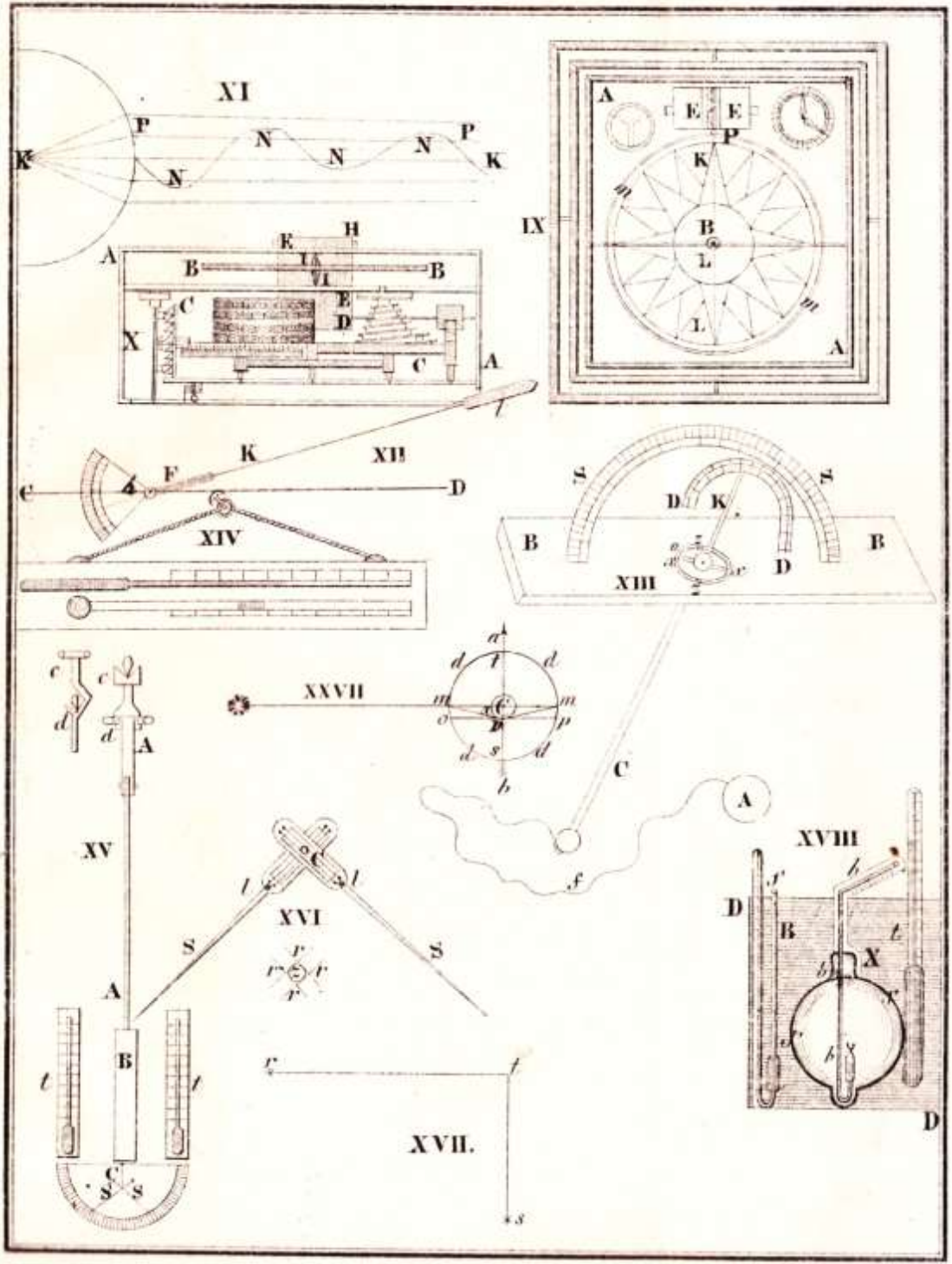

Figs. IX-XVIII, XXVII.

# CHAPTER III
# ABOUT THE MEANS BY WHICH ERRORS IN THE SHIP'S COURSE CAUSED BY THE SEA CURRENT SHOULD BE CORRECTED

§ 48

Anyone can foresee how much hope there is to progress further in the known, as soon as the vast multitude and variety of ocean currents present themselves according to differences in location and time. Great errors are made and will still be corrected! One should expect comfort and assistance from a single learned navigator. Meanwhile, one should not lose heart, but stretch one's thoughts even further, the more desperate the situation seems. It cannot be blamed that in the previous chapter the diligence in calculating sea waves was put forth, while here great distances are left out of the calculation of the ship's path. But we take comfort in the example of astronomers, who, when calculating the motion of planets and fixed stars, are careful even about seconds, while when they study the revolutions of comets, they consider whole years barely for errors.

§ 49

Thus, when the theory of the movement of ocean waters is very imperfect (about which assertion, however, my opinion is not without benefit to the scientific world below this, it seems useful to discuss for sailors), instruments must still be used to intermittently test the sea currents.

§ 50

I do not mention other known methods, which are commonly used for this investigation, but I prefer the one that is based on the following art. That is, that seawater moves faster the closer to the surface, and at the very top the fastest; conversely, at a certain depth, it is completely calm, not feeling the action of forces from the winds or celestial bodies.

§ 51

To determine this, a copper sphere *A* (Fig. XIII) should be thrown into the water, attached to a rope *f* from the stern, tied to spoke *C*, which, connecting, extends with indicator *K*, moving alongside semicircle *SS*, divided into degrees. At the end of the indicator, also attach semicircle *DD*, divided into degrees, perpendicular to *SS*. The entire apparatus should be secured to board *BB,* which should be fastened behind the cabin. The length of the rope should be determined through practice, as well as the size and weight of the sphere. Center *O* should be attached to board *BB* on two axes *xx* and *zz*, so that the spoke with the indicator can freely rotate in all directions.

§ 52

By the known position of the sails, stop the ship stationary on the sea surface, throw ball A into the sea, which the deeper it sinks, the more resistance it will feel from the quiet water below. The rope will tighten, the spoke and indicator will tilt, showing the action along the length of the ship in semicircle *SS*, and along the width on semicircle *DD*. From both inclinations, the speed and direction of the sea current can be conveniently determined, which must first be brought to the required level through experiments.

§ 53

The inclinations of the indicator, depending on the ship's rocking, coming to their extreme limits, must be diligently and attentively noted, then divided in half: the midpoint will show the true inclination of the sea current. This rule should be observed in every use of maritime instruments when the ship is swaying.

CHAPTER IV
**ABOUT THE MEANS OF DISCOVERING AND CORRECTING ERRORS ARISING FROM DIFFERENT COMPASS INCLINATIONS**

§ 54

Drawings for recognizing this inclination on the ocean, composed from observations that are not quite accurate, as much as they can please in use, are well known to those practicing navigation. However, for the lack of exact and unquestionable knowledge, such success is not useless. Meanwhile, proposing some means is not in vain, I think, which, as it seems, can be used by sailors.

§ 55

Of these, the first is nothing but a guess, which in very dark times can provide some comfort, consists in the agreement of magnetic inclination with inclination. Many observations confirm that the closer the inclination of the magnetic needle to the meridian, the deeper it is. Observing this and comparing it with a diagram of magnetic inclinations, one can have some confidence in overcast weather when the sky is covered with clouds everywhere. Another method, although much more accurate and reliable, but without some clarity of the sky, even through small openings in the clouds, cannot be used, consists in the following compass.

§ 56

Circle *dd* (Figs. XXV, XXVI), on which the winds are depicted, must rotate between open jaws *b*, which, by moving other spring jaws c aside, can close, grip the edge of the compass circle, and completely stop its movement, which is done by rudder *f* and lead *g*. The observer must take the instrument by handle *t* and through eyepieces *pp* (which can be inclined to the horizon in various ways) aim at some visible star, or the Moon, or during the day the Sun. And when aiming through both eyepieces, immediately press lead *g* with a finger; at that very moment, the edge of the compass circle will be pinched in the jaws. The time should be shown, giving a signal, and line *rr*, passing parallel to the upper side of the jaws and with the compass diameter, will indicate the degrees by which the magnetic needle deviates from the vertical circle[24] of the observed star; and from there, at the known time on the clock, the magnetic needle's inclination can be determined.

§ 57

This is all that a sailor should use for good during cloudy weather. Let him expect better from the science of navigation, which I briefly consider next.

---

[24] the vertical line on which the star was observed.

Figs. XIX-XXVI.

PART THREE
**ON NAUTICAL SCIENCE**

CHAPTER I
**ABOUT THE NAUTICAL ACADEMY**

§ 58

Navigation — a matter so important — until now has been carried out almost solely through practice. For although academies and schools have been established for the education of maritime affairs with benefit, they only teach what is already known, so that young people, having acquired the necessary skill in this knowledge, replace the elderly, taking their places. As for such institutions, composed of people, in mathematics, especially in astronomy, hydrography, and mechanics, skilled in useful inventions to increase the safety of navigation, no one, as far as I know, has undertaken constant care.

§ 59

Such an academy or assembly could conveniently be established by those who gain great wealth from navigation, so that the maintenance of a certain number of learned individuals forming a society could be considered as nothing against their treasures. Considering the extent of this matter, scholars living in various parts of the world would unite in unity of purpose and each would present their findings to a common authority, from which they would receive support.

§ 60

The function of such an academy would consist of the following: 1) Following the example of assemblies of various travels on land and sea, as abbreviated with praise in England, gather from various books everything that has been written so far in favor of navigation. To search, from wherever possible, for reliable maritime notes; to publish useful selections to the public, so that not only the members of the assembly, but also others could use them to promote safe navigation. 2) To establish by common council what and how should be investigated in the future, and what assistance should be required from those provided. 3) What is most important: to encourage respectable enterprises in navigation with promises of decent rewards and to encourage learned individuals and those capable in this matter. 4) To arrange voyages for learned sailors. But all this should be comprehensively prescribed when establishing specific regulations.

# CHAPTER II
## ON THE COMPOSITION OF A TRUE MAGNETIC THEORY

### § 61

To establish theory from observations, and to correct observations through theory, is the best method for seeking truth. Therefore, in the magnetic theory, the most subtle of all matters in physics, this should be done. From those reflections, which from a few known phenomena almost show magnificent light to the learned world, deductions cannot increase the usefulness of sensitive navigation. For changes in phenomena due to differences in place and time are so diverse that, apart from the most delicate and laborious high mathematics, they drown out almost all the strength of human attention. Here I mention not with disdain the excellent knowledge of algebra, which I regard as the highest degree of human knowledge, but I only reason that it should be used in its place after observations have been collected.

### § 62

A multitude of observations will be the best help in this matter, which are of two kinds. The first consists of experiments conducted by a person who loves to test nature in one place; the second contains records made by sailors without the desired accuracy. The first should be followed initially in testing the causes, and the others should be used with consideration in further investigations, until better ones are found in the future.

### § 63

In such reflections, one should keep in mind that the parts of each magnet differ in strength due to their different qualities, and the same should be thought about the vast earthly body. Not based on assumption, but based on nature.

### § 64

I consider the Earth a magnet[25]: for a magnet is nothing but iron ore, just like the whole terrestrial sphere, because there is almost no species of earth or stone that does not exhibit signs of iron; there is not a single country in the world where there is no iron ore, in which the quality differs from various lands, just as the quality varies in different parts of the magnet.

### § 65

---

[25] the conclusion that the globe is a huge magnet was expressed as early as 1600 by W. Gilbert in his work "De Magnete" (On the Magnet).

Therefore, when other similar magnets, that is, the main bodies of light, especially those closer to it, revolve in its gravitational sphere, then due to the changed position, its magnetic force is confused in various ways, which, due to the different quality of parts of this great magnet, acts differently, and for this reason, the position of the magnetic needle changes in different places and at different times. For if the whole body of the terrestrial sphere were of the same material, the magnetic force would have a consistent effect everywhere in inclination and in the compass deviation over time, or, on the contrary, if the planet's position always remained the same, the magnetic force would differ in direction depending on the place, and not on the time.

§ 66

If anyone wants to see this clearly, let him combine several magnets, poles, and axes in similar positions, so that a magnetic sphere is composed of them. Let him add a special needle to each magnet, indicating inclination and deviation; then, taking a strong special magnet, let him rotate it at a moderate distance from the composite magnetic sphere and from that he will discern what we should think about our terrestrial magnet.

§ 67

My reasoning extends there to arouse the attention of sailors, as well as those traveling around the Earth, to test the magnetic force in all countries accessible to man. For, according to the opinion proposed by my love for art, without many and faithful observations of each place, a general theory about changes in magnetic force cannot be established. For frequent observations, especially made in clear weather on a calm sea, I recommend using a compass with eyepieces, as described by me above (§ 56).

§ 68

However, I do not consider it excessive to conduct magnetic experiments, like those made by Delagyre [26] and others, with spherical magnets made like the Earth, not with the intention of finding an exact correspondence of changes in the magnetic needle around the Earth and around the magnetic sphere, for the different natures of the parts do not allow this thought, but in order to find a general law by which the positions of magnetic spheres change the position of the magnetic needle due to differences in meridians and distances from the equator, and especially in their different positions relative to each other, from which to gain a clearer understanding of the actions of the great terrestrial magnet.

---

[26] Philippe de LaHire (1640-1719) - French mathematician, astronomer, painter and architect.

# CHAPTER III
## ON THE COMPOSITION OF THE THEORY OF MARINE CURRENTS

### § 69

It is well known how many marine movements correspond to the flow of the Moon and the Sun; therefore, no one will dispute that the true theory of sea currents should be sought from here, taking into account the depth of the seas and the shores. Let others attribute this phenomenon to some attraction or pressure; it seems more appropriate to me to consider it as interference in traction according to my theory.

### § 70

When the main celestial bodies move swiftly, they do not carry gravitational matter with them but create a new sphere of influence around them at every point, similar to the spreading sound waves in the air. When a body emitting sound moves rapidly in calm air, it excites the air around it, causing various sounds to be perceived. Just as it is impossible for all the air in which the whistle is made to fly around the arrow that produces the whistle, the sound wave only travels through it. It has the property of achieving its purpose with a single vibration. Similarly, it is impossible to imagine that the sphere of gravitational matter, traveling with tremendous speed alongside a planet, being the outermost layer of fluid, would carry the entire atmosphere with it. Like a magnet, which communicates its force to many iron objects without feeling any loss itself because the liquid matter present everywhere fills the loss in its sphere; just as a stone thrown from a sling does not lose its force as it passes through the air but collects new matter along its path and brings it into orderly motion around itself; just as iron, without touching the magnet, acquires magnetic force that it did not have before; just as light, repelled by a mirror, obeys its movements at an incomprehensible speed, assuming various colors and shapes—similarly, around a moving planet, new gravitational matter must gather at every point of its orbit.

### § 71

Having established this, what consequences do we observe? In the origin of light, it is noted that in its rapid expansion, it somewhat deviates. And this should also be assumed in the assembly around the current planet of gravitational spheres, that it lags somewhat in its performance. From this, the movement of the Earth and other planets around their axes, as well as the flow of the ocean, occurs, as I will show in the following.

### § 72

Let's assume that *ab* (Fig. XXVII) is a part of the circle along which the Earth completes its annual path around the Sun, *dd* is the equator, *mm* is the meridian of the Sun at noon. Lines *mr* from the meridian, where it intersects the equator, extending to point *r*, which is the center of gravity, behind due to the slower movement of the Earth from the geometric Earth center *C* due to the lag in the assembly of the gravitational sphere, *op* represents the intersection of the Earth along the circle that parallelly crosses the equator through point *r*. It follows from this that the line *sr* is shorter than the semidiameter *sC,* and the line *rt* is longer. From the laws of mechanical gravity, it is known that the force of gravity acts in inverse square proportion to the distance from the center of heavy bodies. Therefore, the gravity towards center *r* in *s* is stronger than in *t*. Moreover, from the curvilinear motion of the Earth around the Sun, it is concluded that the gravitational matter towards the Sun S forces the Earth, whence it is evident that the forces exert themselves on the sides of the Earth *s* and *t*. And since forces acting inconsistently cause interference due to their different strength, the gravitational forces towards the Earth's center in *t* and s hinder the force towards the Sun: that is, the force in s hinders more than the force in *t*. Therefore, the gravitational force in *t* towards the Sun acts more powerfully for lesser elevation, and it moves part of the Earth *otp* towards the Sun more quickly than the other part *osp*, which causes the leading part of the Earth *otp* to incline towards the Sun *S*. Meanwhile, the center remaining behind due to the delay in the late assembly of the gravitational sphere moves from *r* to *h*; and thus, half of the Earth, being the leading part along the annual path, always being heavier towards the Sun, inclines towards it and seeks its equilibrium, which it will not find until perhaps its annual course is interrupted.

§ 73

In this case, how much the Moon and other planets in proximity to the Earth disturb the center *r*, I do not discuss for brevity, and many observations are required for this purpose. Therefore, the fact that the Earth's equator is not parallel to the plane of the ecliptic suggests thinking about the inequality of the Earth's sphere itself. Because when we consider that on its midnight half, all of Europe, all of Asia and North America, three-quarters of Africa rise above the sea horizon, whereas on the opposite southern half, only part of South America, and even then not all, includes a quarter of Africa and several islands (unknown lands cannot be so vast as to fill this deficiency, as evidenced by distant voyages in the southern hemisphere), then we can reasonably assume that the center of the Earth's gravity is not coincident with the center to which falling bodies tend, and that the northern hemisphere of the southern hemisphere is heavier, which may cause a shift in the Earth's movement around the axis towards the poles and produce an angle between the ecliptic and the equator.

§ 74

Let's assume that on the back side s, the distance from the center *r* is less than on the front half *t*. Therefore, in this place, all bodies are lighter than in the latter. Hence, it follows that a liquid body, like water, should descend in *s* according to hydrostatic rules, rise higher in *tt*, and rise even higher in *t*, thus it is necessary for the general wave to move towards the front side and to be once a day. The similarity with the general flow of the ocean from east to west and with the tides and ebbs can be judged when observations arranged in the following manner at different locations are made and collected.

§ 75

From the records of the Royal Paris Academy, it is known about the plumb line used to study the changes in the direction of falling bodies towards the center. But as far as I know, this has been left out entirely. Perhaps, for the great length of such an instrument, there was neither the ability nor the occasion, and in shorter distances, such a change would have been difficult to detect. To renew this phenomenon, worthy of attention, I invented a method to establish a plumb line in ordinary rest with a length of many fathoms, which I produced as follows. To the brass strip *A* (Fig. XV, XVI) with a length of a fathom, I attached two poods of lead *B* at the lower end, and I hung it at the upper end on two cushions *cd* so that the plumb line could swing from east to west and from north to south. At the lower end, I fixed a thin cylindrical center *C*, which would move freely in the short ends of the arrows *SS* between double-crossed hairs in such a way that one arrow indicated movement to the east, and the other to the west. The distance from the center in the plumb line to the axes on which the arrows rotate is 3 and 1/2 of linea[27], and the arrows are half a foot long. Hence, it is evident that the length of the plumb line has been increased to seventeen fathoms. To ensure equal warmth on both sides, two thermometers *t, t* were placed.

§ 76

Observing the movements of this great pendulum, I deliberately noted the regular changes that occur more sensitively from east to west than from north to south; a table containing six hundred of my observations is attached[28].

§ 77

Does the increase and decrease in gravity occur from the change in the falling bodies' center? I attempted to test it with this method[29]. I placed an ordinary barometer *bb* (Fig. XVIII) in a glass sphere with a diameter of ten inches. I placed the sphere in vessel *DD* filled with water with ice. Hole *X* was sealed to prevent water from entering the sphere and, in short, to prevent the gravity of the external air, changes in temperature, and cold affecting the air contained inside the sphere and the barometer. Thermometer *t* was used to indicate the constant temperature of the water, and barometer *R* with an open hole *f* above the water for comparing mercury elevations.

---

[27] Linea = 1/10 of an inch; 3 ½ linea = 8.9 mm
[28] See Appendix II.
[29] Reference to the world's first gravimeter (fig. XXV), based on the static principle. The prototype of this gravimeter is the so-called "universal barometer", the design of which Lomonosov proposed in a slightly different form as early as 1749 (PSS [2], vol. 2, pp. 327–337), having also developed a plan of experiments with this instrument (ibid., pp. 339–343). Being a modification of the "universal barometer" and based on the same principle, the instrument described in § 77 (fig. XXV) was the world's first gravimeter, through which Lomonosov conducted observations of changes in gravity force along with prolonged observations of the oscillations of a plumb line under the influence of changes in the tide-generating forces of the Moon and the Sun. The results of these observations are contained in the "Tables of Oscillations of the Centrosophic Pendulum Observed in Petersburg," published in this volume ([2], v.4, pp. 489–708 and 800–816).

From this, I tried to ascertain whether changes in gravity, consistent with the changes observed in the plumb line, followed. Many inconveniences of variable weather, especially the advancing spring, did not allow me to ascertain the true cause of the changes I observed. In the coming winter, by repeating the experiments, I hope to be sure of it and declare it to the scientific community.

### § 78

However, as these experiments require diligent repetition and verification at different locations, I advise all cautious investigators of natural hidden actions to establish such plumb lines in ancient large stone buildings where there is no danger of any deviation from the perpendicular line, which are all the better the longer and heavier the lead weights. The deep cellar of the Paris Observatory is safe from any instability in this case, and especially the mines in Saxony and the Harz Mountains are immensely suitable for this purpose, if the local enthusiasts for science were willing to devote little effort and diligence to it. I do not mention that in India and America, such experiments, greatly serving for this theory, could be promoted by learned individuals and patrons of science.

## CHAPTER IV
## ON THE PREDICTION OF WEATHER, ESPECIALLY WINDS

### § 79

Predicting the weather, if necessary and useful, is better known to the farmer on Earth, to whom during sowing and harvesting, rain, warmed by the sun, is necessary, and to the sailor at sea, who would greatly benefit if he could always point to the side from which long-lasting winds will blow or where a sudden storm will strike.

### § 80

But all this should be expected from the true theory of the movement of liquid bodies around the globe, that is, water and air. Both are subject to the same reasons, except that air, in addition to the variations in general gravity, is also subject to the action of solar rays and the warmth from underground passing through open seas into the atmosphere during winter.

### § 81

I observed and concluded waves in the atmosphere, which, according to the theory explained above (§ 74), should exist in large liquid bodies around the globe[30]. A remarkable agreement; we see it between the hot zone with its constant winds and the relatively stable barometer. Initially, I considered the significant fluctuations in the barometer outside the hot zone to be due to the battles of opposing winds and their dispersal, and that the first would lead to an increase and the latter to a decrease in mercury. However, upon further investigation, I realized that wind battles occur only in the lower atmosphere because significant changes in solar heat occur in it, and according to its magnitude, they should affect the battles of the winds. But it is known that the lower layer of the atmosphere under the hot zone is much higher than in climates lying outside it, so the fluctuations in the barometer would need to be greater, especially since there are many great and strong wind battles there, despite the constancy of the usual eastern winds.

§ 82

Therefore, I consider the main reason for the significant fluctuations in the mercury levels in the atmosphere to be greater here than under the hot zone[31]. For the upper part of the atmosphere, following the influence of the Moon and the Sun, can more quickly traverse a degree of longitude at a given latitude, for example, sixty degrees, than under the equator, due to the magnitude of this being twice as great. Therefore, the air can more quickly gather into waves, rise higher, and load that part of the atmosphere more heavily. And the further north the circles parallel to the equator diminish, the higher the atmospheric waves ascend, and the more sensitively the barometer changes.

§ 83

However, it would be impossible for the orderly flow of these waves to occur due to the absorption of different heat in the air from the Sun and from the earth's depths. All of this, according to the true theory, should be confirmed and arranged into order by frequent and accurate observations and records of atmospheric changes made by sailors. Especially when in different parts of the world, in different states, those who use navigation have established self-recording meteorological observatories, the location and establishment of which with various new instruments I have a new idea about, requiring special description.

§ 84

---

[30] Reference to Lomonosov's wave theory of cyclone genesis. Something similar to atmospheric waves—vortices—was allegedly first noted by Fitzroy and Dove (in 1790–1800, i.e., significantly later than Lomonosov), to whom an priority in the creation of the wave theory was traditionally attributed (seemingly, incorrectly). On Lomonosov's priority in the creation of this theory see S.P. Khromov. *Fundamentals of Synoptic Meteorology,* 1948, p. 5.

[31] Referring here to the middle latitudes, where as a result of wave processes, the so-called polar front is formed, Lomonosov shows: 1) that surface pressure is linked to wave processes in the atmosphere's depth and 2) that wave processes in temperate latitudes are more active than in the tropics (in terms of frequency).

Upon concluding this discussion on weather prediction, I cannot but further delight sailors by providing them with a new marine barometer. It is well known how useful it is to anticipate severe and dangerous storms in advance to avoid unforeseen mishaps. On land, the barometer foretells them several hours, and sometimes even a day in advance, by suddenly falling or rising considerably. An ordinary barometer cannot be used at sea. Therefore, I construct it from two thermometers: one of triple spirit, the other atmospheric, which is especially called a manometer (Fig. XIV). Both are fixed horizontally on one board, first determining the freezing point in water with ice with one, then assigning another limit at about 90 degrees in warm water and dividing everything as necessary. Then, write down the degree of the usual barometer's height at that time. It is known that the first thermometer varies with one change in temperature, while the manometer feels the change in temperature and air pressure. Therefore, when both thermometers move together, indicating the same degree, it means that the barometer stands at the same height as when those two thermometers were made. When the atmospheric thermometer stands lower than the other, it indicates that the air has become heavier and the barometer higher; and when the atmospheric thermometer stands higher than the vodka thermometer, it indicates that the air has become lighter and the barometer lower.

## CONCLUSION

Having reasoned that there are many dangers in the sea, to which not only ships, constructed with great effort and expense and laden with costly goods, are subject, but also human lives, no one will be surprised that those engaged in the sciences seek various means to avert them.

To the salvation of such wealth, all efforts must be exerted against such a great and fearsome monster as the ocean, employing all manner of valor and cunning. Similarly, considering the variety of reasons by which sailors are sometimes led astray from their intended course, no one will regard the variety of instruments as superfluous. For the magnetic force changes with different positions, not corresponding to either the currents of the sea or the winds, the ocean moves differently, regardless of the position of the magnetic needle; the waves fluctuate with different inclinations, neither obeying the magnet's inclination nor the current of the sea, but yielding to the unified breath of the winds. Different things by nature require different tools. And the Creator Himself has provided suitable organs for sight, according to the properties of refracted light, for hearing, according to the ability of vibrating air, and for other senses. Therefore, against such diverse actions or, rather, wars of the unstable sea, all possibilities of reasoning, power, and wealth must be employed. Oh, if only those labors, cares, expenses, and countless multitudes, which war abducts and destroys, were employed in favor of peaceful and scholarly navigation, then not only unknown shores in the inhabited world, not only coasts connected under the icy poles, would be discovered, but even the secrets of the sea floor might have been explored through prudent human endeavor! Through the mutual exchange of excesses, our happiness has grown so much! And the day of learning would shine even brighter with the revelation of new natural mysteries!

We fervently desire and hope for such coveted success in calming the storms of war in Europe[32], through the glorious deeds of Russian heroism! And envisioning the recently celebrated sacred anointing and coronation to our benevolent autocrat of all the Russias[33], as a divine pledge of bounty to us, we cannot think otherwise than that her happiness will multiply and our pleasures on land and sea will be confirmed, and universal joy with her resounding glory will endure throughout eternity, unparalleled.

ADDENDUM I

1. While this reasoning was being printed, I invented a new instrument, which although not large, is satisfied with such advantages for making observations for the accurate determination of latitude and longitude by the Moon at sea, that 1) without any division of the quadrant, it can show the time at the ship's location, as well as latitude and longitude; 2) it eliminates all disturbances in observations from the gloomy horizon, 3) arising from the variable refraction of rays. For the sake of simplicity and smallness, every sailor can purchase it and use it freely. It also consists of two mirrors, as described above. Determining the position of the Moon with fixed stars can also be done by observing the edge of the Moon with them on a vertical circle. Another time is required for the description of this method and for putting it into useful practice.

2. However, I will strive to make separate descriptions for making each instrument proposed in this reasoning, and for the experiments in actual operation with the required tables, as far as possible, to publish each separately.

3. Here I remind the reader that [Fig. XXI], the description of which is omitted in the reasoning itself, depicts an observatory for the training of young maritime observers on land, so that on the curved beams *AA*, crossed and secured, the ship observatory can move like a ship tossed by waves by jerking on ropes *ff*, and so that the observer on land becomes accustomed to overcoming the swaying of the body by the movement of the body on a maritime equilibrium during the very action of sea agitation, which can be attached to the mast *r* [Fig. XXIII] and to the iron rod *s* by the screw *t*.

---

[32] Reference the Seven Years' War (1756-1763), in which Russia participated from 1757 to 1761.
[33] Reference to the celebration of the coronation day of Empress Elizabeth Petrovna, held annually on April 25.

# ADDENDUM II
## OBSERVATIONS OF PENDULUM VARIATIONS, INDICATING THE CENTER TOWARDS WHICH FALLING BODIES TEND.[34]

      Perhaps the reader may suspect that these pendulum variations arise from changes in heat and cold or from building oscillations; however, the first doubt is dispelled by the fact that red-hot coals placed next to one of the thermometers produced, over the course of an hour, a difference of eight degrees between the thermometers, while the plumb line readings changed, the meridian [pendulum] detected a change of 2/10 of a line, and the east one did not show any significant change. And since the usual temperature difference between the thermometers never reached two degrees, it is clear that the change observed on the plumb lines is not dependent on temperature fluctuations. The second doubt is decisively irrelevant, as the building cannot be subject to such changes. For 1) they are periodic and correspond to the movement of the Sun and Moon, which I will publish in a separate treatise as the number of observations increases over time. 2) The changes are greater in the direction of the length of the house, i.e., eastward, and smaller in the direction of its width, i.e., meridional, whereas the reverse should be true, as the length of the building relates to the width as 3 to 1. 3) The southern wall of the building is illuminated by the Sun for 12 hours straight, and currently, in spring, the earth thaws in the southern part earlier than in the northern; therefore, if the building were inclined, it would lean more towards the south, but current observations of the plumb line show the opposite. For the number increases from east to west more than from south to north.

      Furthermore, the same southern part is washed by the Moika River; it is clear that for this reason also, the building should more likely slope towards the south; however, the movement of the plumb line shows the opposite. Hence, any doubt that the pendulum changes arise not from building oscillations but from actual changes in the center of gravity is apparently dispelled.

But diligent continuation of observations and comparison with similar experiments conducted in various places will finally eliminate any difficulties.

---

[34] In the following tables: the first column is the date and time (March and April 1759, letters M. and P. indicate morning (AM) and evening (PM), correspondingly); the second column O.O. gives the reading of the East-West position of the pendulum (Fig. XV); the third column B.A. is for the position along North-South direction. The 80 lbs pendulum is described in § 77 above – it was 2.1m long, and it's E-W and N-S readings were amplified by a factor of 60/3.5=17.14 by the Lomonosov's two-arrow device (Fig.XVI).

## APPENDIX II.
Continens obseruationes directionum penduli, quae ostendunt mutationes centri grauium.

| Martius. | | O.O | B.A. | Martius. | | O.O | B.A. |
|---|---|---|---|---|---|---|---|
| 13 | 4 P. | 2½ + | 90 ⅕ | 18 | 12 P. | 1 ⁶⁄₁₀ | 90 ¼ |
| 14 | 7 M. | 3 . | 90 ⅕ | — | 1 P. | 1 ⁷⁄₁₀ | 90 ¼ |
| — | 9½ M. | 2 ⅝ | 90 ½ | — | 6½ P. | 2 . | 90 ⁴⁄₁₀ |
| — | 1 P. | 2 ⅜ | 90 ⅚ | — | 10 P. | 2 . | 90 ⅝ |
| — | 5 P. | 2 ⅜ | 90 | 19 | 6½ M. | 2 ½ | 90 ⁵⁄₁₀ |
| — | 10 P. | 2 ⅞ | 90 ½ | — | 8 M. | 2 ½ | 90 ⅖ |
| — | 12 P. | 2 ⅞ | 90 ⅝ | — | 10½ M. | 2 ⁴⁄₁₀ | 90 ⁷⁄₁₀ |
| 15 | 7½ M. | 3 ⅛ | 90 ⁵⁄₁₀ | — | 1 P. | 1 ⁹⁄₁₀ | 90 ⅗ |
| — | 9 M. | 3 ⅚ | 90 ⅓ | — | 5½ P. | 2 . | 90 ⅖ |
| — | 2 P. | 2 ⅝ | 90 ⅓ | — | 8 P. | 2 . | 90 ¼ |
| — | 6 P. | 2 ⅔ | 90 ⁰⁄₁₀ | — | 9½ P. | 2 ½ | 90 ½ + |
| 16 | 6 M. | 3 ⁸⁄₁₅ | 90 ⁵⁄₁₀ | 20 | 5 M. | 2 ⁶⁄₁₀ | — |
| — | 9½ M. | 3 . | 90 ⁶⁄₁₀ + | — | 6 M. | 2 ⅖ | 90 ⁴⁄₁₀ |
| — | 1⅓ P. | 2 ⅔ | 90 ⁶⁄₁₀ | — | 7½ M. | 2 ⁶⁄₁₀ | 90 ⁴⁄₁₀ |
| — | 4½ P. | 2 ⁴⁄₁₀ | 90 ½ | — | 10¼ M. | 2 ⁴⁄₁₀ | 90 ⁴⁄₁₀ |
| — | 11½ P. | 2 ⁶⁄₁₀ | 90 ⁶⁄₁₀ | — | 12 M. | 2 ⁵⁄₁₀ | 90 ½ |
| 17 | 6 M. | 2 ⅝ + | 90 ⁵⁄₁₀ | — | 5 P. | 2 ⁷⁄₁₀ | 90 ⅝ |
| — | 7 M. | 2 ¼ | 90 ½ | — | 9 P. | 2 ⅗ | 90 ⅜ |
| — | 9 M. | 2 ⅕ | 90 ⅖ | — | 11 P. | 2 ⁴⁄₁₀ | 90 ¼ |
| — | 11 M. | 2 . | 90 ⅓ | 21 | 6½ M. | 2 ⁵⁄₁₀ | 90 ²⁄₁₀ |
| — | 12 M. | 1 ⁷⁄₁₀ | 90 ⅖ | — | 8 M. | 2 ⁵⁄₁₀ | 90 ²⁄₁₀ |
| — | 1½ P. | 1 ¾ | 90 | — | 4 P. | 2 ⁴⁄₁₀ | 90 ⅜ + |
| — | 4 P. | 1 ⁷⁄₁₀ | 90 ⁶⁄₁₀ | — | 7 P. | 2 ⁶⁄₁₀ | 90 ⅗ |
| — | 6 P. | 1 ⁷⁄₁₀ | 90 ⁴⁄₁₀ | — | 10 P. | 2 ⅖ | 90 ⅜ |
| — | 12 P. | 2 ⁵⁄₁₀ | 90 ½ | 22 | 6 M. | 3 . | 90 ¼ |
| 18 | 4 M. | 2 ⁵⁄₁₅ | 90 ⅖ | — | 7 M. | 3 + | 90 ¼ + |
| — | 5½ M. | 2 ⁵⁄₁₀ | 90 ⅓ | — | 10 M. | 2 ⁵⁄₁₀ | 90 ⁷⁄₁₀ |
| — | 7½ M. | 2 ⅔ | 90 ⅜ | — | 1 P. | 2 ⁴⁄₁₀ | 90 ⁷⁄₁₀ |
| — | 10½ M. | 2 . | 90 | — | 5 P. | 2 ⁷⁄₁₀ + | 90 ⅞ + |

| Martius | | O. O. | B. A. | Martius | | O. O. | B. A. |
|---|---|---|---|---|---|---|---|
| 22 | $8\frac{1}{2}$ P. | $2\frac{1}{15}$ | $90\frac{1}{15}$ | 26 | 11 P. | $2\frac{8}{10}$ | $89\frac{8}{10}$ |
| — | $10\frac{3}{4}$ P. | $2\frac{3}{10}$ | $90\frac{1}{10}+$ | 27 | $5\frac{1}{2}$ M. | 3 . | $89\frac{8}{10}$ |
| 23 | $6\frac{1}{4}$ M. | $2\frac{2}{10}+$ | $90\frac{2}{15}$ | — | 8 M. | $3\frac{1}{10}$ | $89\frac{9}{10}$ |
| — | 8 M. | $2\frac{3}{10}+$ | $90\frac{2}{10}$ | — | 10 M. | $3 . +$ | idem |
| — | $9\frac{1}{2}$ M. | $2\frac{1}{10}$ | $90\frac{2}{10}+$ | — | $2\frac{3}{4}$ P. | $2\frac{8}{10}$ | idem |
| — | 1 P. | $2\frac{2}{10}$ | $90\frac{3}{10}$ | — | 6 P. | $3 +$ | idem |
| — | 6 P. | $2\frac{1}{10}$ | $90\frac{2}{10}$ | — | 9 P. | $3\frac{1}{10}$ | idem |
| — | 10 P. | 2 . | $90\frac{2}{10}$ | — | 11 P. | $3\frac{2}{10}$ | 90 — |
| 24 | 5 M. | $2\frac{1}{2}$ | $90 +$ | — | 12 P. | idem | 90 — |
| — | 6 M. | $2\frac{1}{4}+$ | $90\frac{1}{10}$ | 28 | $5\frac{3}{4}$ M. | $3\frac{2}{10}+$ | $89\frac{9}{10}$ |
| — | 7 M. | $2\frac{1}{2}+$ | $90\frac{1}{10}+$ | — | 7 M. | $3\frac{1}{4}$ | $89\frac{8}{10}$ |
| — | $9\frac{1}{2}$ M. | $2\frac{1}{2}$ | $90\frac{1}{10}—$ | — | 9 M. | $3\frac{2}{10}—$ | $89\frac{8}{10}$ |
| — | 12 M. | $2 +$ | 90 . | — | $1\frac{1}{2}$ P. | $2\frac{6}{10}$ | $89\frac{8}{10}$ |
| — | 5 P. | $1\frac{7}{10}+$ | $90 +$ | — | 3 P. | $2\frac{1}{2}+$ | idem |
| — | 7 P. | 2 . | $90 +$ | — | 5 P. | $2\frac{6}{10}$ | idem |
| — | 10 P. | $2\frac{2}{10}$ | 90 | — | 9 P. | $2\frac{8}{10}$ | idem |
| 25 | 6 M. | $2\frac{1}{2}$ | 90 — | — | 11 P. | $2\frac{9}{10}$ | idem + |
| — | $7\frac{1}{2}$ M. | $2\frac{1}{2}+$ | 90 — | 29 | 5 M. | $3\frac{1}{4}$ | $89\frac{7}{10}$ |
| — | 10 M. | $2\frac{1}{2}$ | $90 +$ | — | 7 M. | $3\frac{1}{4}$ | $89\frac{3}{4}$ |
| — | 1 P. | $2\frac{2}{10}+$ | $90\frac{1}{10}$ | — | $10\frac{1}{2}$ M. | 3 . | $89\frac{9}{10}$ |
| — | 3 P. | $2\frac{3}{10}$ | $90\frac{1}{10}$ | — | $6\frac{1}{2}$ P. | $2\frac{1}{2}$ | $89\frac{9}{10}$ |
| — | 4 P. | $2\frac{3}{10}$ | idem | 30 | 6 M. | $3\frac{8}{10}+$ | $89\frac{9}{10}$ |
| — | 5 P. | $2\frac{3}{10}$ | idem | — | 8 M. | $3\frac{2}{10}$ | idem |
| — | 6 P. | $2\frac{1}{2}$ | $90\frac{1}{10}$ | — | $10\frac{1}{2}$ M. | $3\frac{1}{10}+$ | idem |
| — | 7 P. | $2\frac{1}{2}$ | idem | — | 7 P. | $3\frac{4}{10}$ | idem |
| — | 8 P. | $2\frac{1}{2}$ | 90 . | — | $11\frac{1}{2}$ P. | $3\frac{4}{10}$ | idem — |
| — | 11 P. | $2\frac{2}{10}—$ | $90 +$ | 31 | 6 M. | $3\frac{3}{4}$ | $89\frac{3}{4}$ |
| 26 | $6\frac{1}{2}$ M. | 3 . | 90 . | — | $8\frac{1}{4}$ M. | $3\frac{6}{10}+$ | idem |
| — | 11 M. | $2\frac{6}{10}$ | $89\frac{9}{10}$ | — | 10 M. | $3\frac{5}{10}$ | $89\frac{6}{10}$ |
| — | 5 P. | $2\frac{3}{4}$ | idem + | — | 1 P. | $3\frac{3}{10}$ | $89\frac{1}{2}$ |
| — | 6 P. | $2\frac{9}{10}$ | $89\frac{8}{10}$ | — | 3 . P. | $3\frac{3}{10}$ | $89\frac{1}{2}+$ |

| Martius. | O.O. | B. A. | Aprilis. | O O | B A |
|---|---|---|---|---|---|
| 31 \| 5½ P. | 3 4/15 | 89 5/15 | 6 \| 6¾ P. | 3 7/10 | 89 4/10 |
| — 7 P. | 3 ½ | 89 0/15 | — 9¼ P. | 3 ½ | idem |
| — 8½ P. | 3 6/10 | 89 0/10 | 7 \| 6 M. | 4 1/15 + | 89 4/10 |
| — 10 P. | 3 6/10 + | 89 6/15 | — 7½ M | 4 . | 89 3/10 |
| Aprilis. | | | — 12 M. | 3 6/15 + | 89 ¼ |
| | | | — 2 P. | 3 7/10 | 89 3/15 |
| 1 \| 2 M. | 3 7/10 | 89 6/15 | — 4 P. | 3 7/15 | 89 3/15 |
| — 4¾ M. | 3 8/15 | idem | — 6 P. | 3 8/15 | 89 3/15 |
| — 7 M. | 3 8/15 | idem | — 9⅓ P. | 3 7/15 | 89 ¼ |
| — 9 M. | 3 ¾ | idem | 8 \| 5¼ M. | 3 7/15 | 89 3/15 |
| — 10¼ M. | 3 7/10 | 89 6/15 | — 7¼ M. | 3 6/15 + | idem |
| — 1½ P. | 3 4/15 | 89 ¼ + | — 12 M | idem | 89 1/10 |
| — 6¼ P. | 3 4/10 + | 89 ¼ | — 3 P. | 3 6/10 | 89 3/10 + |
| — 7¼ P. | 3 ⅛ | 89 6/15 | — 6 P. | 3 7/10 | 89 2/10 |
| — 10 P. | 3 ⅜ + | idem | — 9 P. | 3 ¼ | 89 1/10 |
| 2 \| 5 M. | 4 — | idem | 9 \| 4½ M. | 3 6/15 — | 89 1/10 — |
| 3 \| 12 M. | 3 ⅜ | 89 6/15 | — 6 M | 3 6/15 | 89 1/7 |
| 4 \| 6½ M | 4 2/15 | idem | — 1 P. | 3 7/15 | 89 1/10 |
| — 10 M. | 4 8/15 | idem | — 2½ P. | 7 ¾ | 89 2/15 |
| — 11½ M. | 4 — | idem | — 7¼ P. | 3 8/15 | 89 1/10 |
| — 4 P. | 3 8/15 | 89 7/10 + | — 10 P. | 3 ⅛ | 89 2/15 |
| — 8½ P. | 4 — | 89 7/10 | — 11 P. | 3 4/10 | 89 1/15 |
| 5 \| 5 M | 4 ¼ | idem | 10 \| 4½ M. | 3 ⅛ | 89 1/10 |
| — 8 M. | 4 1/15 | 89 6/10 — | — 2¼ P. | 3 1/10 | 89 1/10 |
| — 1 P. | 3 7/15 | 89 ½ | — 5 P. | 3 4/15 | 89 2/10 |
| — 3 P. | 3 7/10 — | 89 ⅔ + | — 6 P | 3 4/10 | 89 2/10 + |
| — 6 P. | 3 ⅜ | 89 6/10 | — 8½ P. | 3 6/10 | 89 1/10 + |
| — 9 P. | 3 ⅔ | 89 ½ + | 11 \| 1 M | 3 8/10 | 89 2/10 |
| 6 \| 6 M. | 4 ⅛ | 89 4/15 | — 8 M. | 3 4/10 + | 89 2/10 + |
| — 9¼ M | 3 6/15 | 89 4/10 + | — 10¼ M. | 3 4/15 | idem |
| — 11¾ M. | 3 ½ + | 89 ½ — | — 2½ P. | 3 7/10 | 89 1/15 |

| Aprilis. | | | O.O. | B.A. | Aprilis. | | | O.O. | B.A. |
|---|---|---|---|---|---|---|---|---|---|
| 11 | 5 | P. | $3\frac{1}{2}$ | $89\frac{2}{10}$ | 16 | 1 | P. | $4\frac{4}{10}$ | $89\frac{1}{2}$ |
|  | $9\frac{1}{4}$ | P. | $3\frac{2}{10}$ | $89\frac{2}{10}$ |  | 3 | P. | $4\frac{4}{10}-$ | $89\frac{1}{2}-$ |
| 12 | $5\frac{1}{2}$ | M. | $4\frac{1}{10}-$ | $89\frac{1}{4}$ |  | $4\frac{3}{4}$ | P | idem | idem |
|  | 7 | M. | $4\frac{1}{10}$ | idem |  | 6 | P. | $4\frac{6}{10}$ | $89\frac{1}{2}+$ |
|  | $9\frac{3}{4}$ | M. | $4\frac{3}{10}+$ | $89\frac{3}{10}$ |  | 7 | P. | idem | idem |
|  | $3\frac{1}{2}$ | P. | $4 \;\; +$ | $89\frac{4}{10}$ |  | $9\frac{3}{4}$ | P. | $4\frac{4}{10}$ | $89\frac{1}{2}$ |
|  | 6 | P. | $3\frac{8}{10}$ | $89\frac{4}{10}$ | 17 | $4\frac{3}{4}$ | M. | $4\frac{6}{10}$ | $89\frac{3}{10}$ |
|  | 7 | P. | $3\frac{5}{10}$ | $89\frac{3}{10}+$ |  | 8 | M. | $4\frac{1}{2}$ | $89\frac{3}{10}$ |
|  | $10\frac{1}{2}$ | P. | $3\frac{4}{10}$ | $89\frac{2}{10}$ |  | 9 | M | $4\frac{3}{10}$ | idem |
| 13 | 5 | M. | $3\frac{2}{10}$ | idem |  | 10 | M. | $4\frac{3}{10}-$ | $89\frac{4}{10}-$ |
|  | $6\frac{1}{2}$ | M. | $3\frac{0}{10}$ | idem |  | 11 | M. | $4\frac{1}{2}$ | idem |
|  | $8\frac{1}{2}$ | M. | $3\frac{0}{10}-$ | idem |  | 12 | M. | idem | idem |
|  | 12 | M. | $3\frac{2}{10}$ | $89\frac{1}{4}$ |  | 3 | P. | $4\frac{1}{4}$ | idem |
|  | 4 | P. | $3\frac{3}{10}$ |  |  | 4 | P. | $4\frac{2}{10}$ | $89\frac{4}{10}+$ |
|  | $7\frac{1}{2}$ | P. | $3\frac{1}{2}+$ | $89\frac{3}{10}$ |  | 5 | P. | $4\frac{1}{4}$ | idem |
|  | 9 | P. | idem | idem |  | 8 | P. | $4\frac{1}{4}$ | idem |
| 14 | $7\frac{3}{4}$ | M | $3\frac{9}{10}$ | $89\frac{1}{4}$ |  | 9 | P | $4\frac{1}{2}-$ | idem |
|  | 9 | M. | $4 \;\; -$ | $89\frac{3}{10}$ | 18 | $4\frac{3}{4}$ | M. | $4\frac{7}{10}$ | $89\frac{4}{10}+$ |
|  | 10 | M. | $4 \;\; .$ | $89\frac{1}{4}$ |  | 6 | M. | $4\frac{3}{4}$ | $89\frac{1}{2}$ |
|  | 12 | M. | $4 \;\; . +$ | $89\frac{3}{10}$ |  | $7\frac{1}{2}$ | M | idem | $89\frac{1}{2}-$ |
|  | 6 | P. | $4 \;\; . +$ | $89\frac{4}{10}$ |  | 9 | M. | $4\frac{7}{10}$ | idem |
|  | 10 | P | $4\frac{2}{10}$ | idem |  | 10 | M. | $4\frac{6}{10}+$ | idem |
| 15 | 6 | M. | $4\frac{4}{10}$ | idem |  | $12\frac{1}{2}$ |  | $4\frac{6}{10}$ | $89\frac{1}{2}$ |
|  | 9 | M. | $4\frac{3}{10}$ | $89\frac{4}{10}$ |  | 3 | P. | idem | $89\frac{1}{2}-$ |
|  | 1 | P. | $4\frac{3}{10}$ | $89\frac{4}{10}+$ | 19 | 8 | M. | $4\frac{3}{10}$ | $89\frac{4}{10}$ |
|  | 4 | P | $4\frac{3}{10}$ | idem |  | 3 | P. | $4\frac{6}{10}$ | $89\frac{3}{10}$ |
|  | 7 | P | $4\frac{4}{10}+$ | $89\frac{1}{2}$ |  | 7 | P. | $4\frac{7}{10}$ | $89\frac{1}{2}$ |
| 16 | 6 | M | $4\frac{3}{4}+$ | $89\frac{4}{10}+$ | 20 | 7 | M. | $5 \;\; .$ | $89\frac{1}{2}+$ |
|  | 7 | M. | $4\frac{1}{2}+$ | $89\frac{1}{4}$ |  | 1 | P. | $4\frac{2}{10}$ | $89\frac{4}{10}$ |
|  | 8 | M | $4\frac{1}{2}$ | $89\frac{1}{2}-$ | 27 | 6 | M. | $4\frac{2}{10}$ | $89\frac{1}{4}$ |
|  | 9 | M. | $4\frac{1}{2}-$ | idem |  | 7 | M. | $4\frac{7}{10}$ | $89\frac{2}{10}$ |

| Aprilis. | | | O. O. | B. A. | Aprilis. | | | O. O. | B. A. |
|---|---|---|---|---|---|---|---|---|---|
| 27 | 9 | M. | idem | idem | 28 | 9 | P. | $4\frac{6}{10}$ — | idem |
| — | 11 | M. | $4\frac{6}{10}$ + | $89\frac{2}{10}$ + | — | 10 | P. | $4\frac{6}{10}$ — | idem |
| — | 12 | M. | $4\frac{6}{10}$ | idem | 29 | $5\frac{1}{2}$ | M. | 5 — | $89\frac{1}{10}$ |
| — | $3\frac{1}{2}$ | P. | $4\frac{1}{2}$ | $89\frac{1}{4}$ | — | 7 | M. | 5 . | $89\frac{1}{10}$ — |
| — | 6 | P. | $4\frac{1}{2}$ | $89\frac{1}{4}$ | — | $8\frac{1}{2}$ | M. | $4\frac{9}{10}$ | $89\frac{1}{10}$ |
| — | 7 | P. | $4\frac{1}{2}$ — | idem | — | 11 | M. | $4\frac{8}{10}$ — | $89\frac{1}{10}$ — |
| — | 9 | P. | idem | $80\frac{3}{10}$ | — | $1\frac{1}{2}$ | M. | $4\frac{8}{10}$ | idem |
| — | $10\frac{1}{2}$ | P. | $4\frac{6}{10}$ — | $89\frac{3}{10}$ | — | 4 | P. | $4\frac{7}{10}$ | idem |
| 28 | $4\frac{3}{4}$ | M. | $4\frac{7}{10}$ | $89\frac{1}{2}$ | — | 6 | P. | $4\frac{1}{2}$ | $89\frac{1}{10}$ + |
| — | 6 | M. | $4\frac{8}{10}$ | $89\frac{1}{4}$ | — | $8\frac{1}{2}$ — | P. | $4\frac{1}{2}$ | $89\frac{2}{10}$ |
| — | 7 | M. | $4\frac{7}{10}$ | $89\frac{1}{10}$ | — | 10 | P. | $4\frac{6}{10}$ | $89\frac{2}{10}$ |
| — | $8\frac{1}{4}$ | M | $4\frac{3}{4}$ | $89\frac{1}{4}$ | 30 | 5 | M. | 5 . | $89\frac{2}{10}$ |
| — | $10\frac{1}{2}$ | M. | $4\frac{7}{10}$ | $89\frac{1}{4}$ | — | 6 | M | 5 . | $89\frac{2}{10}$ — |
| — | 12 | M. | $4\frac{6}{10}$ | $89\frac{2}{10}$ | — | 7 | M | idem | idem |
| — | 2 | P. | $4\frac{6}{10}$ | $89\frac{1}{10}$ + | — | 9 | M. | $4\frac{6}{10}$ | idem |
| — | 3 | P. | idem | $89\frac{2}{10}$ | — | 11 | M. | $4\frac{8}{10}$ + | idem |
| — | 4 | P. | $4\frac{6}{10}$ | $89\frac{2}{10}$ — | — | $1\frac{1}{2}$ | P | $4\frac{6}{10}$ | idem |
| — | 5 | P. | $4\frac{6}{10}$ | $89\frac{2}{10}$ | — | 5 | P. | $4\frac{6}{10}$ + | $89\frac{2}{10}$ + |
| — | $6\frac{1}{2}$ | P. | $4\frac{6}{10}$ | $89\frac{2}{10}$ | — | 12 | P. | $4\frac{6}{10}$ + | $89\frac{2}{10}$ |
| — | 8 | P. | $4\frac{1}{2}$ — | $89\frac{2}{10}$ — | | | | | |

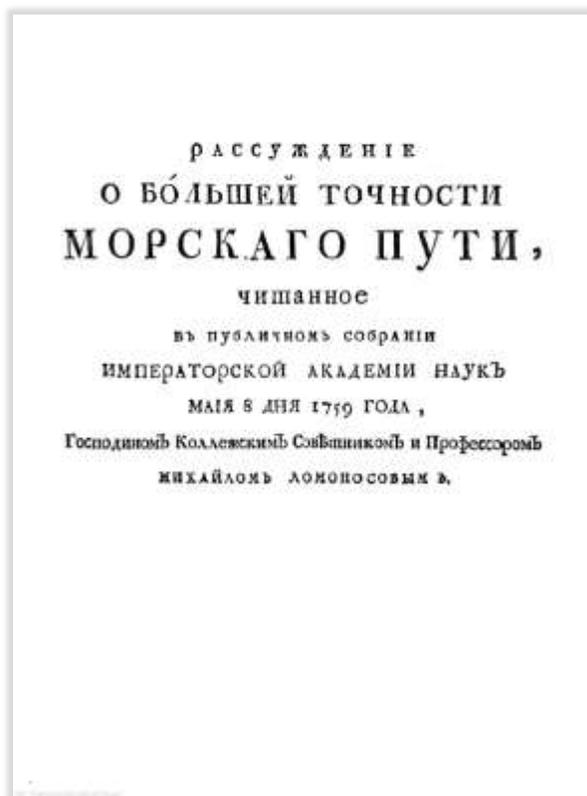 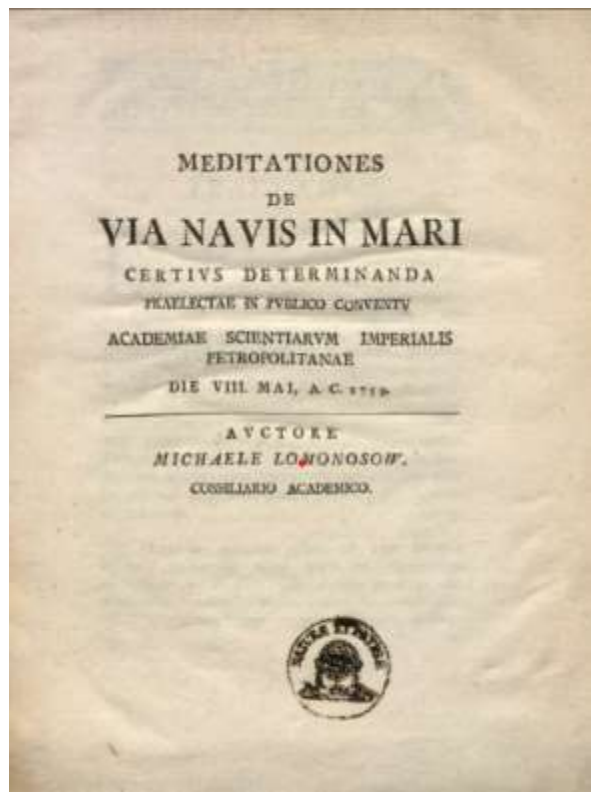

*Commentary* [by V.Shiltsev]:

This English translation of Mikhail Lomonosov's seminal work draws from its original Russian and Latin sources [1] (also Ref. [2], v.4, pp.123-319.). It concludes the series of English translations of Lomonosov's nine most significant scientific treatises, all of which Lomonosov himself compiled in the volume *titled Lomonosow Opera Academica*, intended for dissemination among European Academies. Other translations of these works can be found in Refs. [3-9]. This rendition was initially composed with the aid of Google Translate and ChatGPT programs, providing a rough draft from both Russian and Latin versions of Lomonosov's "*Discourse…*". However, machine translations are found to be not fully satisfactory, as the meaning of the Russian original was sometimes lost or distorted due to the difficult 18th-century Russian of Lomonosov's original publication, which may not always be easily readable or understood even by contemporary Russian speakers. They often failed to capture the nuances and complexities of the original text, and therefore, extensive reworking and improvements by the translator were indispensable to ensure the translation accurately conveys the intended meaning and is free from semantic deficiencies. For further exploration of the life and contributions of Mikhail Lomonosov (1711-1765) - the eminent Russian polymath and a towering figure of the European Enlightenment - readers are encouraged to consult various books [10-12] and recent articles [13-20].

This "*Discourse*..." represents a speech prepared by Lomonosov for presentation at a public session of the Imperial Academy of Sciences on May 8, 1759. Between February and April of that year, the original Russian text was crafted, later translated by the author into Latin, accompanied by substantial augmentations and revisions. The inaugural publication of the "*Discourse*..." occurred as separate editions in Russian and Latin during 1759. During the public assembly, Lomonosov shared only a fraction of his prepared address, focusing on "his innovative contributions to the science of navigation and the associated machinery."

In the introduction, Lomonosov underscores the significance of navigation for humanity, highlighting its peril due to the imprecise determination of a ship's position. Traditional methods involve determining latitude through the measurement of celestial bodies' altitude and longitude by comparing local time with that of the prime meridian. Alternatively, one can track a ship's coordinates on a map based on its initial location, course, and speed. Lomonosov details the numerous challenges inherent in both approaches and proposes in this "*Discourse*..." his invented navigation tools alongside a discourse on the theory of ocean currents and the magnetic compass. The subsequent portion of the work delves into "scientific navigation," wherein Lomonosov proposes enhancements to existing instruments such as the "English quadrant," spring, and hourglass clocks, as well as innovations in observation platform suspension for minimized ship motion interference. He further elucidates methods for ascertaining local time, latitude, and the prime meridian using celestial bodies.

In the second part of the "Discourse...," novel methodologies for determining ship coordinates in cloudy weather are explored, leveraging instruments like a large self-writing compass, a clinometer for crosswind deviation, and a mechanical log line ("dromometer") for continual speed measurement. Lomonosov introduces a simple mechanical device, the salometer, to gauge the influence of ocean currents on ship direction and speed. The third section underscores the importance of "scientific navigation," advocating for international collaboration through the establishment of a "navigation academy" and proposing systematic observations to develop a comprehensive understanding of magnetic declination.

A special place in the "*Discourse*..." is occupied by the chapter "On the Compilation of the Theory of Ocean Currents" in which Lomonosov presents his hypothesis of the origin of tides, the cause of which is not the attraction of the Moon (he, a consistent adherent of the mechanistic view of the world, did not accept long-distance action), but the daily change in the force of gravity on the Earth's surface, due to the fact that the geometric center of the Earth does not coincide with the center of gravity, moving during the day along a circular trajectory. Where the center of gravity is closer to the Earth's surface, there is an increase in the force of terrestrial attraction and ebb tide, and vice versa. Lomonosov also describes his 7 ft long "centroscopic pendulum" installed in the basement of his house, and the results of his multi-year observations, which, as he believed, confirmed his theory. Then he describes the design of a mercury barometer, with which he hoped to observe changes in gravity throughout the day and confirm the readings of the "centroscopic pendulum." Appendix I discusses Lomonosov's invention of a simplified tool for determining time and ship coordinates at sea in fair weather, along with a "trainer" for simulating ship movement on land during navigation exercises. Appendix II provides "Tables of Oscillations of the Centroscopic Pendulum" alongside explanatory notes.

I would like to thank Prof. Robert Crease of SUNY, my long-term collaborator and co-author of several scholar papers on Mikhail Lomonosov, for the encouragement to translate Lomonosov's major works to English.